\documentclass{aa}
\usepackage[varg]{txfonts}
\usepackage{graphicx}
\usepackage[version=3]{mhchem} 
\usepackage{longtable}
\bibpunct{(}{)}{;}{a}{}{,} 
\usepackage{color}

\newcommand{\IRAS}{IRAS~16293$-$2422}
\newcommand{\iras}{IRAS~16293}

\begin{document}
\title{A cold accretion flow onto one component of a multiple protostellar system}

\author{N. M. Murillo\inst{1,2} \and E. F. van Dishoeck\inst{2,3} \and A. Hacar\inst{4,2} \and D. Harsono\inst{5} \and J. K. J{\o}rgensen\inst{6}}

\institute{Star and Planet Formation Laboratory, RIKEN Cluster for Pioneering Research, Wako, Saitama 351-0198, Japan\\ \email{nadia.murillomejias@riken.jp}
	    \and Leiden Observatory, Leiden University, Niels Bohrweg 2, 2300 RA, Leiden, the Netherlands
        \and Max-Planck-Institut f\"{u}r extraterrestrische Physik, Giessenbachstra\ss e 1, 85748, Garching bei M\"{u}nchen, Germany
        \and Department of Astrophysics (IfA), University of Vienna, T\"{u}rkenschanzstra\ss e 17, 1180, Vienna, Austria
        \and Institute of Astronomy and Astrophysics, Academia Sinica, No. 1, Sec. 4, Roosevelt Road, Taipei 10617, Taiwan, R. O. C.
	    \and Niels Bohr Institute, University of Copenhagen, {\O}ster Voldgade 5–7, DK-1350 Copenhagen K., Denmark}

\abstract
{Gas accretion flows transport material from the cloud core onto the protostar. In multiple protostellar systems, it is not clear if the delivery mechanism is preferential or more evenly distributed among the components.}
{The distribution of gas accretion flows within the cloud core of the deeply embedded, chemically rich, low-mass multiple protostellar system \IRAS~is explored out to 6000 AU.}
{Atacama Large Millimeter/submillimeter Array (ALMA) Band 3 observations of low-$J$ transitions of various molecules such as \ce{HNC}, cyanopolyynes (\ce{HC3N}, \ce{HC5N}), and \ce{N2H+} are used to probe the cloud core structure of \IRAS~at $\sim$100 AU resolution. Additional Band 3 archival data provide low-$J$ \ce{HCN} and \ce{SiO} lines. These data are compared with the corresponding higher-$J$ lines from the PILS Band 7 data for excitation analysis. The \ce{HNC}/\ce{HCN} ratio is used as a temperature tracer.}
{The low-$J$ transitions of \ce{HC3N}, \ce{HC5N}, \ce{HNC} and \ce{N2H+} trace extended and elongated structures from 6000 AU down to $\sim$100 AU, without accompanying dust continuum emission. Two structures are identified: one traces a flow that is likely accreting toward the most luminous component of the system \IRAS~A. Temperatures inferred from the \ce{HCN}/\ce{HNC} ratio suggest that the gas in this flow is cold, between 10 and 30 K. The other structure is part of an UV-irradiated cavity wall entrained by one of the outflows driven by the source. The two outflows driven by \IRAS~A present different molecular gas distributions.}
{Accretion of cold gas is seen from 6000 AU scales onto \IRAS~A, but not onto source B, indicating that cloud core material accretion is competitive due to feedback onto a dominant component in an embedded multiple protostellar system. The preferential delivery of material could explain the higher luminosity and multiplicity of source A compared to source B. The results of this work demonstrate that several different molecular species, and multiple transitions of each species, are needed to confirm and characterize accretion flows in protostellar cloud cores.} 


\titlerunning{Extended structure in \IRAS}
\authorrunning{Murillo et al.}

\maketitle

\section{Introduction}
Protostellar cloud cores provide the bulk of material that will eventually be accreted onto the forming protostar and disk \citep{li2014}.
The distribution of material is a key factor in understanding how protostars form and evolve.
However, it is not yet well understood how material is delivered from the cloud core onto the protostar and disk.
In particular, it remains unclear how cloud core material is distributed among the individual components of multiple protostellar systems, and how that distribution affects the properties of the resulting system.

Early theoretical work proposed self-similar spherical collapse of the protostellar cloud core \citep{ebert1955,bonnor1956,hayashi1966,larson1969,penston1969,bondi1952,shu1977}.
Hydrodynamical simulations of star formation, on the other hand, suggest that infalling streams of gas, so-called accretion flows, spiral arms or arc-like structures, form as material is accreted onto the protostar \citep{offner2010,seifried2014,li2014,matsumoto2015}.
These gas structures are further shaped by fragmentation and the characteristics of the resulting multiple protostellar system (e.g., \citealt{bate1997,vorobyov2010,young2015,matsumoto2019,kuffmeier2019}).
Such structures are known to be common on parsec scales for high-mass sources (e.g., \citealt{peretto2013}). 
However, limited data are available for lower mass sources due to lack of sensitivity to the material in such streams with low surface brightness.

Observations in the last decade have started to reveal extended structures at a range of scales. 
In embedded protostellar systems, these structures have been traced in continuum at scales below 100 AU (e.g., \citealt{alves2019}), from scales of a few 100 to a few 1000 AU in \ce{HCO+} \citep{tokuda2014}, or \ce{CO} \citep{takakuwa2017,tokuda2018,alves2020}, and in \ce{HC3N} at scales of 10000 AU \citep{pineda2020}. 
These detections are for close (separations $<$ 100 AU) binary systems \citep{takakuwa2017,alves2019,pineda2020} or single protostellar sources \citep{tokuda2014,tokuda2018,alves2020}.
Furthermore, an extended structure in dust and gas, a so-called tail structure, is observed toward a Class II system \citep{akiyama2019,ginski2021}.
The connection between the structures at different scales is not yet clear.
In parallel, the interpretation of these structures varies: from structures created by turbulent fragmentation or shocks \citep{tokuda2014,tokuda2018}, interaction with the environment \citep{akiyama2019}, accretion flows \citep{alves2019,pineda2020,alves2020,ginski2021}, to structures within the protostellar disk \citep{takakuwa2017}.
While some of these observations provide evidence for accretion flows or streamers in low-mass embedded protostellar systems, it is not yet clear how material from such streams is connected to the individual components in wide (separations larger than disk radius) multiple protostellar systems.

The difficulty of detecting cloud core structure stems from the physical conditions of the gas and dust in the cloud core.
In hydrodynamical simulations \citep{li2013,kuffmeier2019} extended structures are present in the dust continuum. 
However in closed-box simulations of cloud cores \citep{harsono2015}, extended structures are not the strongest features in \ce{CO} due to freeze-out.
\ce{CO} isotopologues, common gas tracers in star formation observations, freeze out at temperatures below 30 K and above densities where the freeze-out timescales become shorter than the lifetime of the core \citep{jorgensen2005}.
The presence of outflows, also typically traced in \ce{CO}, further complicates identifying accretion structures, unless their spatial distributions are clearly distinguishable (e.g., \citealt{alves2020}).
At molecular cloud scales, filamentary structures are generally traced in continuum (e.g., \citealt{palmerin2013}; \citealt{wang2020,arzoumanian2021}),  \ce{N2H+} (e.g., \citealt{arzoumanian2013,hacar2013,fernandez2014,hacar2017}), \ce{NH3} (e.g., \citealt{pineda2011,monsch2018,sokolov2019}), and in \ce{HCN}, \ce{HNC}, and \ce{HCO+} (e.g., \citealt{kirk2013,storm2014}; \citealt{hacar2020}).
Given that \ce{N2H+} is formed when \ce{CO} is frozen out while \ce{N2} is present in the gas phase, and is a dense gas tracer \citep{bergin2007}, \ce{N2H+} might also be a good tracer for cold gas structures in the cloud core.
However, processes within cloud cores, such as episodic accretion (e.g., \citealt{hsieh2018}), warm inner envelopes \citep{jorgensen2005,tobin2010,belloche2020}, or gas temperatures below \ce{N2} freeze-out (e.g., VLA1623: \citealt{bergman2011,murillo2018}) can change the distribution of molecular gas along structures in the cloud core.
Cloud cores of both single and multiple protostellar systems present \ce{HCN} emission in single dish observations (e.g., \citealt{murillo2018c}), suggesting it can also be a good tracer of cloud core structure \citep{tafalla2002,tafalla2004}.
Being able to trace how material is transported from molecular cloud to protostellar cloud core scales can shed light onto the mass accretion process in star formation.
However, doing so requires multiple molecular tracers.

\begin{figure*}
	\includegraphics[width=\textwidth]{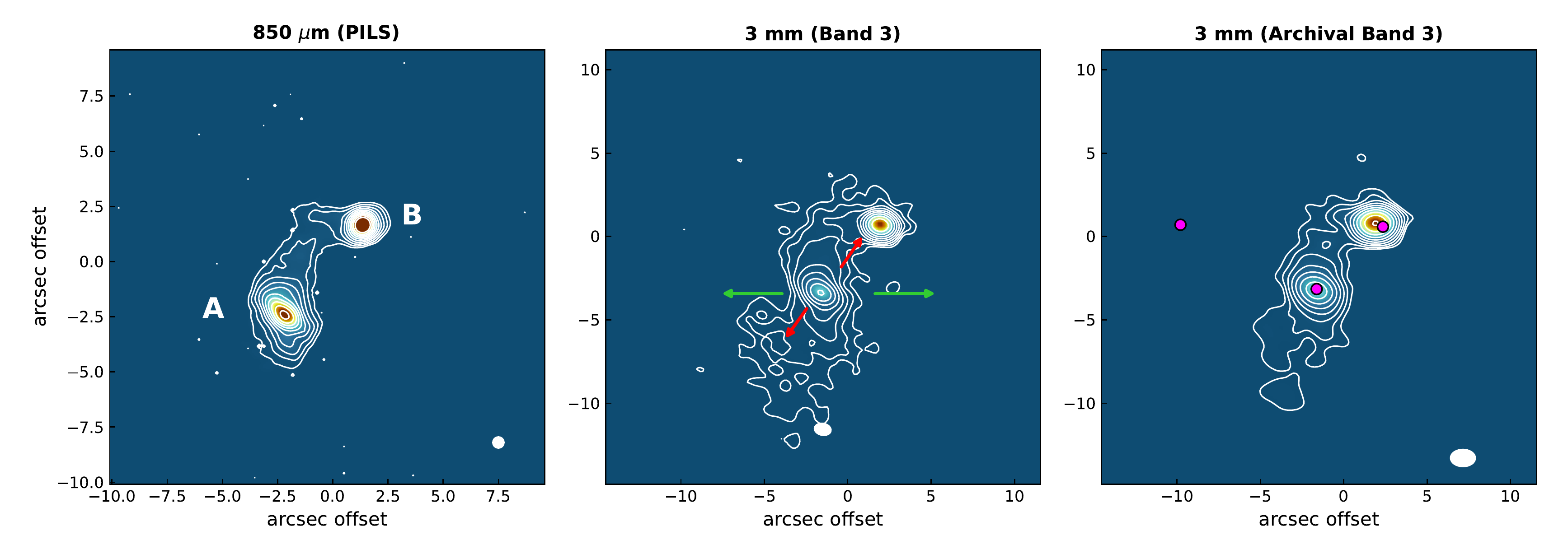}
	\caption{\IRAS~ observed with ALMA. Dust continuum observations at 870 $\mu$m (PILS, \textit{left}) and 3 mm (\textit{center and right}). The levels are spaced logarithmically from 30 to 900~mJy~beam$^{-1}$ for 870$\mu$m, 0.6 to 170~mJy~beam$^{-1}$ for 3mm, and 2.7 to 190~mJy~beam$^{-1}$ for the archival 3 mm continuum. The ellipse at bottom right corner shows the synthesized beam. The arrows indicate the east-west (E-W, green) and northwest-southeast (NW-SE, red) outflow directions from source A. The magenta filled circles in the right panel indicate the positions where the spectra shown in Fig.~\ref{fig:spectra} are extracted. The dust continuum emission at both wavelengths shows the bridge of material between the two sources. At 3 mm, the dust continuum extends southeast of source A, likely tracing part of the outflow cavity.}
	\label{fig:continuum}
\end{figure*}

We aim to characterize the cloud core structure of the deeply embedded protostellar system \IRAS~(hereafter \iras) from 6000 AU down to 100 AU scales using Atacama Large Millimeter/submillimeter Array (ALMA) observations in Band 3 (3 mm).
Located in L1689 N \citep{jorgensen2016}, within the $\rho$ Ophiuchus star forming region, at a distance of 147.3 $\pm$ 3.4 pc \citep{ortiz2017}, \iras~drives a complex and extensive system of two outflows \citep{stark2004,yeh2008,vanderWiel2019}. 
At scales of a few 1000 AU (\citealt{vanderWiel2019} and references therein), one outflow is observed to be directed east-west (E-W outflow), with a second outflow orientated northwest to southeast (NW-SE outflow).
The southern source, \iras~A, presents a flattened disk-like structure.
The radius of the disk-like structure is reported to be about 230 AU from molecular gas observations \citep{oya2016}, while gas and dust radiative transfer or non-LTE modeling suggests a radius of 150 AU \citep{jacobsen2018}.
Recent observations and orbital motion analysis confirmed two protostellar sources toward the position of source A \citep{maureira2020}, thus making \iras~a triple protostellar system.
Separated by 720 AU ($\sim$5$\arcsec$) from \iras~A to the north-west, \iras~B is directed pole-on with respect to the line of sight \citep{pineda2012,jorgensen2016}.
Sources A and B have $v_{\rm lsr}$ = 3.1 and 2.7 km~s$^{-1}$, respectively \citep{jorgensen2011}.
To the east of the \iras~cloud core lies the starless core \IRAS~E (hereafter \iras~E), separated by a distance of 90$\arcsec$ \citep{mizuno1990,stark2004}. 
Most studies of \iras~at a few 1000 AU scales have focused on the outflows, or chemical characterization of sources A and B \citep{mizuno1990,ewine1995,narayanan1998,ceccarelli2000,schoier2002,stark2004,jaber2017}.
However, the flow of material from cloud to disk ($<$100 AU) has not been previously addressed at high enough spatial resolution.

In this paper we present spectrally and spatially resolved observations of low-$J$ lines \ce{HNC} 1--0, \ce{HC3N} 10--9, \ce{HC5N} 35--34, \ce{N2H+} 1--0, \ce{HCN} 1--0, \ce{H^{13}CO+} 1--0 and \ce{SiO} 2--1 toward \iras. 
These molecular tracers recover structures on scales from $\sim$150 up to 6000 AU that have not been seen in any previous observations of this source.
The observations used in this paper are detailed in Sect.~\ref{sec:obs}.
The spatial distribution of each molecule treated in this work, as well as the continuum dust emission, are described in Sect.~\ref{sec:results}.
Section~\ref{sec:analysis} aims to characterize the structures traced by the different molecules. 
The results of this work are then placed in context with previous observations of envelope structure in Sect.~\ref{sec:discussion}.
Section~\ref{sec:conclusions} summarizes the results found in this work.

\begin{figure*}
	\centering
	\includegraphics[width=0.68\linewidth]{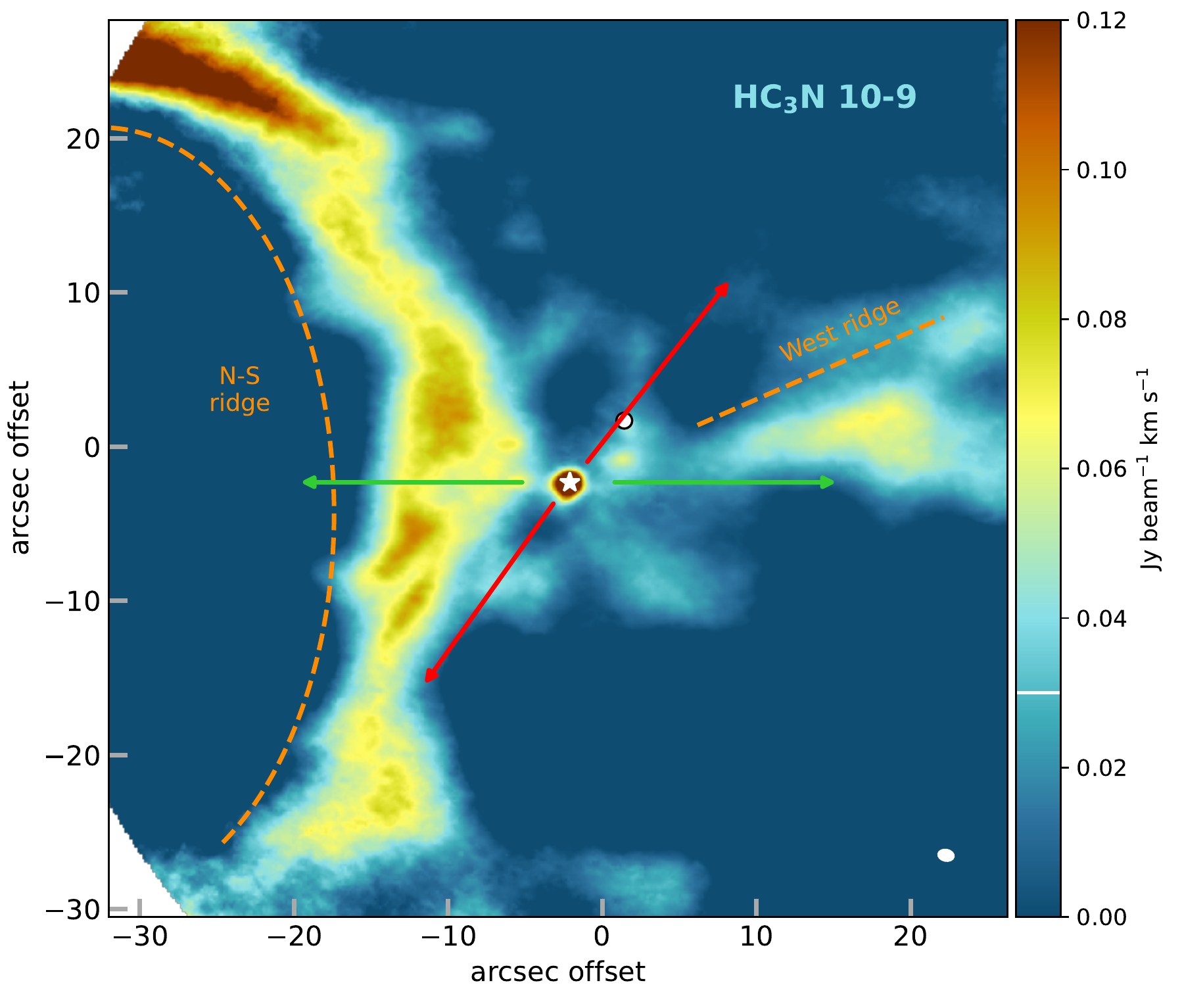}
	\caption{ALMA Band 3 intensity integrated map of \ce{HC3N} 10--9. The 3$\sigma$ level is indicated on the colorbar with a horizontal white line, with $\sigma$ = 0.01 Jy~beam$^{-1}$ km~s$^{-1}$. The positions of \iras~A and B are marked with a white star and filled circle, respectively. The orange dashed lines highlight the structures identified in this work. The arrows indicate the east-west (E-W, green) and northwest-southeast (NW-SE, red) outflow directions from source A. The white ellipse on the bottom right indicates the synthesized beam.}
	\label{fig:HC3N}
\end{figure*}

\begin{table*} 
	\centering 
	\caption{Summary of line observations.}
	\label{tab:lines}
	\begin{tabular}{c c c c c c}
		\hline \hline 
		& & & & & \\
		Line & Transition & $\nu$ & log$_{10}$ $A_{\rm ij}$ & $E_{\rm up}$ & Beamsize \\
		& & (GHz) & (s$^{-1}$) & (K) & arcsec ($\arcsec$) \\
		\hline  
		\multicolumn{6}{c}{Band 3} \\
		\hline
		\ce{HNC} & 1--0 & 90.66356 & -4.57 & 4.4  & 0.95$\times$0.7 \\
		\ce{HCN} & 1--0 & 88.63160 & -4.62 & 4.3 & 1.34$\times$0.94 \\
		\ce{HC3N} & 10--9 & 90.97902 & -4.23 & 24.0 & 0.95$\times$0.7 \\
		\ce{HC5N} & 34--33 & 90.52589 & -4.10 & 76.0 & 0.95$\times$0.7 \\
		\ce{HC5N} & 35--34 & 93.18812 & -4.10 & 80.5  & 0.95$\times$0.7 \\
		\ce{N2H+} & 1--0 $F_{\rm 1}$ = 1--1 & 93.17188 & -4.41 & 4.5 & 0.95$\times$0.7 \\
		& 1--0 $F_{\rm 1}$ = 2--1 & 93.17370 & -4.41 & 4.5 & 0.95$\times$0.7 \\
		& 1--0 $F_{\rm 1}$ = 0--1 & 93.17613 & -4.41 & 4.5 & 0.95$\times$0.7 \\
		\ce{SiO} & 2--1 & 86.84696 & -4.53 & 6.3 & 1.35$\times$0.97 \\
		\ce{H^{13}CO+} & 1--0 & 86.75429 & -4.4 & 4.2 & 0.95$\times$0.7 \\
		\hline
		\multicolumn{6}{c}{Band 7: PILS} \\
		\hline
		\ce{HNC} & 4--3 & 362.63030 & -2.64 & 43.5 & 0.5 \\
		\ce{HCN} & 4--3 & 354.50548 & -2.69 & 42.5 & 0.5 \\
		\ce{HC3N} & 37--36 & 336.52008 & -2.52 & 306 & 0.5 \\
		& 38--37 & 345.60901 & -2.48 & 323 & 0.5 \\
		& 39--38 & 354.69746 & -2.45 & 340 & 0.5 \\
		\ce{H^{13}CO+} & 4--3 & 346.99835 & -2.48 & 41.6 & 0.5 \\
		\hline
	\end{tabular}
	\\
	\tablebib{All rest frequencies were taken from the Cologne Database for Molecular Spectroscopy (CDMS) \citep{CDMS_2016}, and the Jet Propulsion Laboratory (JPL) molecular spectroscopy catalogue \citep{JPL-catalog_1998}. 
		The data for \ce{HNC} are based on \citet{creswell1976} and \citet{thorwirth2000}. The entry for \ce{HCN} is based on \citet{delucia1969} and \citet{delucia1977}.
		The \ce{N2H+} entries are based on \citet{cations_rot_2012}.}
\end{table*}

\section{Observations}
\label{sec:obs}

\subsection{Band 3}
\label{subsec:B3}

ALMA Band 3 observations of \iras~(Project-ID: 2017.1.00518.S; PI: E. F. van Dishoeck) were performed using the 12 m main array.
Observations were carried out with the C43-1 (short baselines) and C43-4 (long baselines) configurations.
The first were carried out on 15 July 2018, and the latter on 27 March 2018.
The $(u,v)$ coverage of the short baselines, 15 to 780 m, recovers structures up to 39$\arcsec$ with a beamsize of 3$\arcsec~\times$2.2$\arcsec$. 
The long baselines, 15 to 1261 m, recover extended material up to 21$\arcsec$ with a beamsize of 0.9$\arcsec~\times$0.7$\arcsec$.
Band 3 provides a field of view of 62$\arcsec$.

The phase center of the observations was located at $\alpha_{J2000}$~=~16:32:22.0; $\delta_{J2000}$~=~$-$24:28:34.0, about 8$\arcsec$ to the west from source B. 
Total on-source times were 85.7 and 171.7 minutes for the short and long baseline configurations, respectively.
The precipitable water vapor (pwv) was in the range of 2.55 to 2.95 mm for the short baselines, and 0.93 to 2.04 mm for the long baselines.
For the short baselines, the quasars J1517--2422, J1427--4206, and J1625--2527 were used for bandpass, flux, and phase, respectively. 
Bandpass and flux calibration for the long baselines was done with J1517--2422, while phase calibration was performed using J1625--2527.

The spectral set-up consists of four spectral windows covering the frequency ranges of: 90.47 to 91.41 (continuum window), 93.14 to 93.20, and 103.28 to 103.51 GHz. 
The last frequency range is distributed in two spectral windows. 
The spectral resolution was of 488.28, 30.52, 61.04 kHz, respectively, corresponding to velocity resolutions of 0.805, 0.049, and 0.088 km~s$^{-1}$. 
Since line widths are typically 3 and 1 km~s$^{-1}$ (FWHM) in source A and B, respectively, not all lines are spectrally resolved.
Molecular line emission was detected in all four spectral windows, consistent with expectations for such a chemically rich source.
This work focuses mainly on the off-source emission from the 90 and 93 GHz spectral windows. 

\subsection{Archival data}
\label{subsec:archB3}

ALMA archival Band 3 observations of \iras~from the project 2015.1.01193.S (PI: V. M. Rivilla) are included in this work. 
Observations were carried out on 11 June 2016, with a phase center of $\alpha_{J2000}$~=~16:32:22.62; $\delta_{J2000}$~=~-24:28:32.46.
The spectral set-up consists of four spectral windows with frequency ranges of: 86.58 to 87.05, 88.50 to 88.97, 99.00 to 99.23, and 101.10 to 101.57 GHz. 
This work focuses on the first and second spectral windows, which contain \ce{HCN}, \ce{SiO} and \ce{H^{13}CO+}.
Both spectral windows have spectral resolution of 488.28 kHz, corresponding to 0.83 km~s$^{-1}$.
\citet{rivilla2019} reported on the data calibration, and observations of the complex organic molecules, but did not include \ce{HCN}, \ce{SiO} and \ce{H^{13}CO+}. 

Total integration time was 70 minutes in the C36-3 configuration with a pwv of 2.0 to 2.4 mm. 
The $(u,v)$ coverage, 15 to 783 m, recovers structures up to 19$\arcsec$, providing a beam of about 1.41$\arcsec~\times$1.02$\arcsec$. 
Bandpass, phase, and flux calibration were performed with the sources J1517--2422, J1625--2527, and Titan, respectively. 

\subsection{Data calibration and imaging}
\label{subsec:datacal}

Calibration was carried out with the \textit{Common Astronomy Software Applications} (CASA) package pipeline \citep{mcmullin2007} for both our Band 3 observations, and the archival Band 3 data.
Self-calibration was performed on both datasets.
Our Band 3 datasets (C43-1 and C43-4) were concatenated prior to self-calibration using the CASA task \textit{concat}.  Self-calibration was performed on the dust continuum emission using the channels provided by the pipeline with additional manual corrections. Phase calibrations were done three times, with \textit{solint} = 32min, 8min, and inf. These corrections were interpolated across the different spectral windows. 

Since the CASA task \textit{uvcontsub} overestimates the continuum subtraction, in particular for line rich observations, a different method was applied.
The self-calibrated data were imaged without continuum subtraction.
The CASA task \textit{tclean} was used to image the datacubes.
Both dust continuum maps and molecular emission channel maps were made using Briggs weighting, cell size of 0.16$\arcsec$, and an image size of 784 pixels. 
Our Band 3 data were all convolved with a beam size of 0.95$\arcsec~\times$0.7$\arcsec$ with a position angle of 79$^{\circ}$ using the \textit{restorebeam} parameter in \textit{tclean}.
Similarly, the archival Band 3 data were convolved with a beam size of 1.41$\arcsec~\times$1.02$\arcsec$ with a position angle of 90$^{\circ}$.
The continuum map was made using a robust parameter of 2.0, while molecular emission channel maps were done with a robust parameter of 0.5.
All images were primary beam corrected with \textit{pblimit} = 0.2.
When producing a full channel map of the non-continuum spectral windows, the parameter \textit{chanchunks} = 8 was used to avoid a core dump due to CASA attempting to use all available memory.
Continuum subtraction was then performed on the imaged channel maps using the python-based tool STATCONT \citep{sanchez2018}.
This tool statistically determines the continuum level from the intensity distribution of spectral data.
Finally, intensity integrated maps were generated from the self calibrated, imaged, and continuum subtracted data.
Maps presented in this work are shown the PILS phase center as the common offset. No regriding of the maps was done unless otherwise specified.
Typical noise levels for molecular line channel maps range between 3 and 20 mJy~beam$^{-1}$. For the intensity integrated maps, noise levels range from 2 to 24 mJy~beam$^{-1}$ km~s$^{-1}$.
The dust continuum noise levels are 0.2 and 0.8 mJy~beam$^{-1}$ for our data and the archival data, respectively. 

\subsection{PILS data}
\label{subsec:PILS}

Data from the Protostellar Interferometric Line Survey (PILS) program (Project-ID: 2013.1.00278.S; PI: Jes K. J{\o}rgensen;
\citealt{jorgensen2016}) is used in this work. 
PILS is an ALMA Cycle 2 unbiased spectral survey in Band 7, using both the 12m array and the Atacama Compact Array (ACA).
The spectral set-up covers a frequency range from 329.147 GHz to 362.896 GHz, and provides a velocity resolution of 0.2 km~s$^{-1}$.
The phase center was $\alpha_{J2000}$~=~16:32:22.72; $\delta_{J2000}$~=~$-$24:28:34.3, set to be equidistant from the two sources A and B at $v_{\rm lsr}$ = 3.1 and 2.7 km~s$^{-1}$ \citep{jorgensen2011}, respectively. 
The resulting $(u,v)$ coverage of the combined 12m array and the ACA observations are sensitive to the distribution of material with an extent of up to 13$\arcsec$ and a circular synthesized beam of 0.5$\arcsec$. 
A detailed description of the observations, reduction and imaging is given in \cite{jorgensen2016}.

\begin{figure*}
	\centering
	\includegraphics[width=0.9\linewidth]{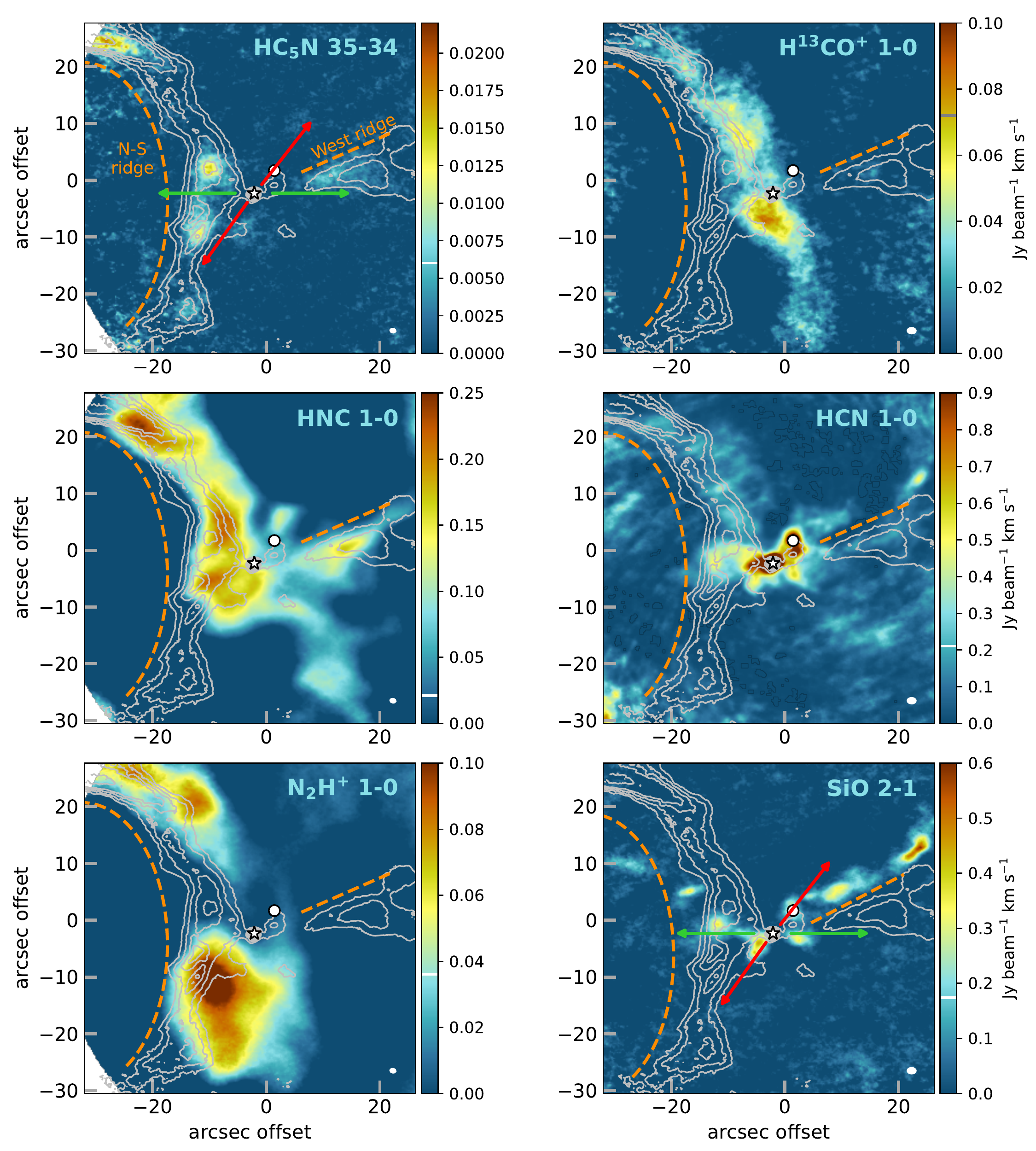}
	\centering
	\caption{ALMA Band 3 intensity integrated maps of the molecular species showing extended emission around \IRAS. The 3$\sigma$ level is indicated on the colorbar with a horizontal white or gray line. \ce{HC3N} 10--9 is overlaid in contours with steps of 3, 5, 7, 9 and 15$\sigma$, with $\sigma$ = 10 mJy~beam$^{-1}$ km~s$^{-1}$. The positions of \iras~A and B are marked with a star and circle, respectively. The orange dashed lines indicate the structures identified in this work. The arrows indicate the east-west (E-W, green) and northwest-southeast (NW-SE, red) outflow directions from source A. The white ellipses on the bottom right indicate the beam of the observations.}
	\label{fig:N2Hp_HC5N_HNC}
\end{figure*}

\section{Results}
\label{sec:results}

\subsection{Continuum}
\label{subsec:cont}
\iras~A and B are detected as separate continuum peaks (Fig.~\ref{fig:continuum}). 
The dust bridge connecting sources A and B, observed at 850 $\mu$m \citep{jorgensen2016,jacobsen2018}, is also detected in 3 mm dust continuum.
Apart from the bridge and sources, the dust continuum emission extends slightly to the southeast of source A.
This is more prominent in the 3 mm observations than in the 850 $\mu$m continuum (Fig.~\ref{fig:continuum}), and is consistent with previous observations (e.g., \citealt{jorgensen2011,jorgensen2016,oya2018,sadavoy2018}).
The direction of the extended dust continuum is consistent with the southeast outflow lobe, and thus probably arises from the outflow cavity.
No additional structures are detected in dust emission beyond the two continuum peaks and the dust bridge, even at this deep level.

\subsection{Molecular emission}
\label{subsec:molemission}

Band 3 spectra were examined at three locations, shown in Figure~\ref{fig:continuum}: on source A, $\sim$1$\arcsec$ west from source B, and an off-source position 12$\arcsec$ east of source B and away from the outflow directions. 
Extracted spectra is shown in Figure~\ref{fig:spectra}.
Seven molecular species are detected at the off-source position: \ce{H^{13}CO+} 1--0, \ce{SiO} 2--1, \ce{HCN} 1--0, two transitions of \ce{HC5N} 34--33 and 35--34\footnote{The \ce{HC5N} 34--33 transition is only detected at the peak locations of the 35--34 transition.
	The difference is most likely due to peak brightness and spectral resolution, both of which are lower for the 34--33 transition (Fig.~\ref{fig:spectra}).
	Thus, only the \ce{HC5N} 35--34 transition is shown in the intensity integrated maps presented in this work.}, \ce{HNC} 1--0, \ce{HC3N} 10--9, and \ce{N2H+} 1--0.
Parameters of each molecular species are listed in Table~\ref{tab:lines}.
Carbon chains, such as \ce{c-C3H2} at 93.16 GHz (log$_{10}$ $A_{ij}$ = -4.76, $E_{\rm up}$ = 92 K), are not detected at this off-source position. 
This is surprising given that several transitions of \ce{c-C3H2} (log$_{10}$ $A_{ij}$ = -2.61 -- -3.23, $E_{\rm up}$ = 38.6 -- 96.49 K) were found to trace extended structure related to the outflow in Band 7 observations (\citealt{murillo2018}; see also Sect.~\ref{subsec:scene}).

\ce{HC3N} 10--9 reveals an extended and elongated ridge of material spanning 60$\arcsec$ ($\sim$8000 AU at 147 pc) stretching from north to south (hereafter N-S ridge), and bending close to the east side of source A (Fig.~\ref{fig:HC3N}). 
This N-S ridge is not seen in dust continuum.
Based on the outflow directions, the south part of the N-S ridge, as well as the region closest to source A, are outflow contaminated (Fig.~\ref{fig:HC3N}).
In contrast, the northern part of the N-S ridge does not coincide with any of the two outflows.
A second, shorter ridge structure extends 20$\arcsec$ west of source A.
This \ce{HC3N} ridge is referred to as the west ridge in the rest of this work (Fig.~\ref{fig:HC3N}).
The west ridge coincides with the north cavity wall of the west outflow lobe \citep{vanderWiel2019}, suggesting it is part of the east-west outflow structure driven by source A.

Intensity integrated maps of the other six molecular species showing extended emission are presented in Figure~\ref{fig:N2Hp_HC5N_HNC}.
Out of the six species, \ce{HC5N}, \ce{HNC}, \ce{N2H+}, \ce{HCN} and \ce{H^{13}CO+} show extended emission along the N-S ridge, with the latter two species presenting only weak emission ($\sim$3$\sigma$).
Both cyanopolyynes are spatially correlated, with \ce{HC5N} and \ce{HC3N} peaking at the same locations along the N-S ridge (Fig.~\ref{fig:N2Hp_HC5N_HNC}).
In contrast, \ce{HNC}, \ce{N2H+}, \ce{H^{13}CO+} and \ce{HCN} are offset to the west relative to the N-S ridge traced in \ce{HC3N}, producing a layered structure.
\ce{N2H+} peaks at $\sim$30$\arcsec$ ($\sim$4500 AU) to the northeast and $\sim$10$\arcsec$ ($\sim$1500 AU) to the southeast from source A, where temperatures are expected to be $\lesssim$20 K (see Section~\ref{subsec:ratios}; \citealt{crimier2010,jacobsen2018}).
Thus, \ce{N2H+} provides a constraint on the temperature along the N-S ridge.
However, \ce{N2H+} presents strong absorption due to filtering, preventing further detailed kinematic and chemical analysis of its emission.

In addition to \ce{HC3N}, the west ridge is detected in \ce{HC5N}, and \ce{HNC} (Fig.~\ref{fig:N2Hp_HC5N_HNC}).
While \ce{HCN} presents some diffuse emission along the west ridge, it is below 3$\sigma$.
The \ce{HCN} blobs show even separations in the northeast to southwest direction, which suggest that the blobs are artifacts of the uv-coverage, rather than real structure.
It is interesting to note that the isomers \ce{HNC} and \ce{HCN} have different peak distributions.
\ce{HCN} is concentrated around the outflow lobes near sources A and B, while \ce{HNC} is more prominent along the N-S and west ridges. 
Comparing the spatial distribution with the direction of both outflows indicates that \ce{HCN} is mainly tracing warmer gas associated to the outflow and its cavity wall. 

On the other hand, \ce{SiO} does not present emission related to either the N-S or west ridges.
Instead, \ce{SiO} traces a string of knots located north of the west ridge, but does not coincide spatially.
The \ce{SiO} clumps around sources A and B are consistent with Band 7 \ce{SiO} observations \citep{vanderWiel2019}.
The brightest \ce{SiO} clump, located 20$\arcsec$ west of \iras~along the string of knots, shows a counterpart in \ce{HCN}.

\section{Analysis}
\label{sec:analysis}

\subsection{Kinematics}


\begin{figure*}
	\centering
	\includegraphics[width=0.9\linewidth]{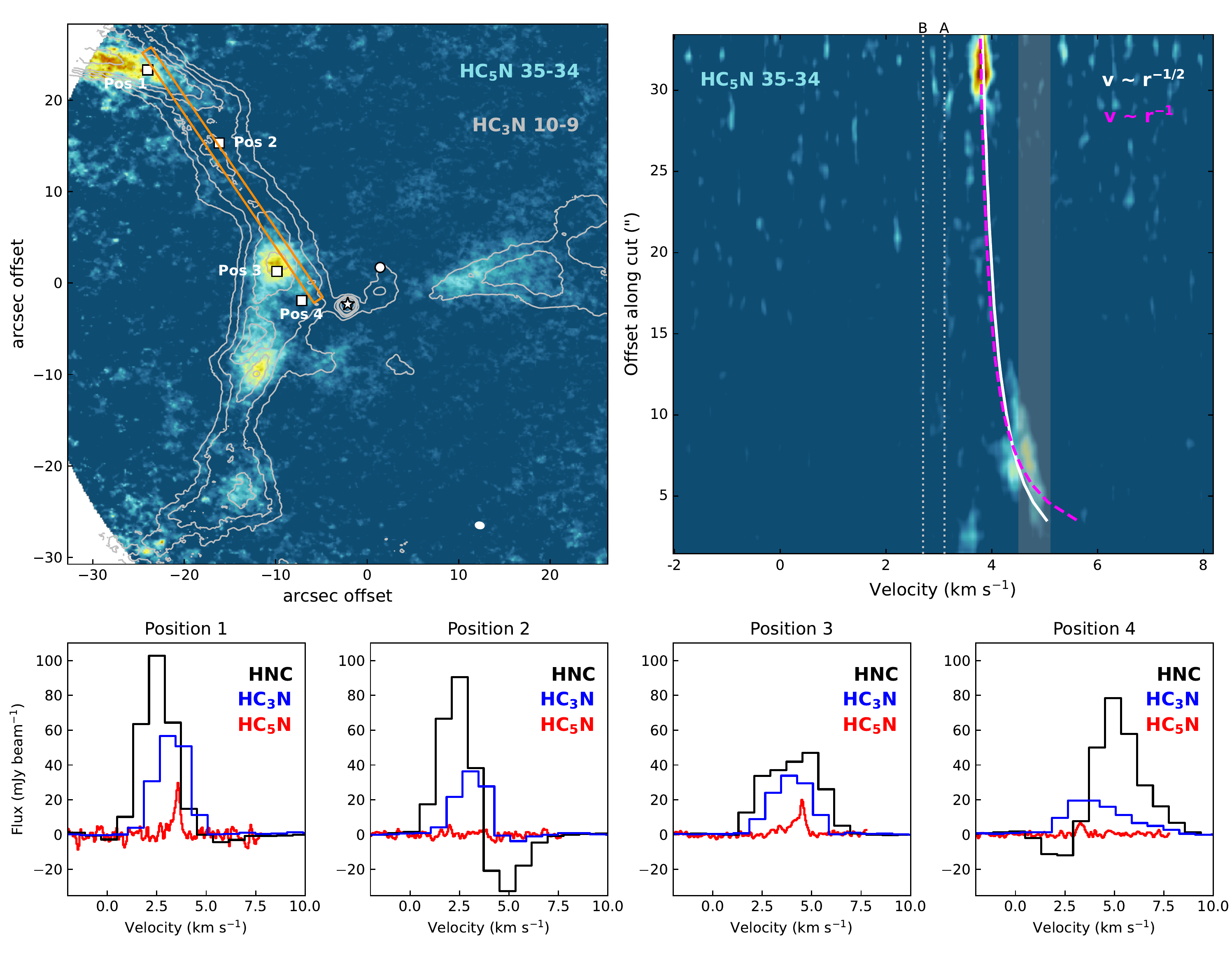}
	\caption{Kinematic analysis of the N-S ridge. Integrated intensity map of \ce{HC5N} (colorscale) overlaid with \ce{HC3N} contours as in Fig.~\ref{fig:N2Hp_HC5N_HNC} (\textit{top left panel}). The white star and circle mark sources A and B, respectively. The white ellipse on the bottom right shows the beamsize. The orange rectangle indicates the cut used to make the position-velocity (PV) diagram  with source A located at offset 0 (\textit{top right panel}). The vertical dotted lines mark the systemic velocity of A and B. The white and magenta lines show free fall and infall curves, respectively. Velocity range with E-W outflow related emission is shaded area (4.5 and 5.1 km~s$^{-1}$) (Fig.~\ref{fig:moment0_outenv}). \ce{HNC}, \ce{HC3N}, and \ce{HC5N} spectra (\textit{bottom row}) were extracted from the corresponding positions marked in the top left panel. The change in peak velocity and line shape are signatures of moving gas. \ce{HNC} and \ce{HC3N} show an inverse P Cygni profile along the middle of the N-S ridge (Position 2, $\sim$20$\arcsec$ from source A), indicating infall. Near source A (Position 4, $\sim$5$\arcsec$) \ce{HNC} shifts to a P Cygni profile, consistent with the presence of outflow in this region.}
	\label{fig:PVdiag}
\end{figure*}

In order to understand the nature of the N-S and west ridges, the gas velocity distributions are examined.
The north part of the N-S ridge is present only in the velocity range between 1.3 and 4.3 km~s$^{-1}$, while the west ridge is detected in the velocity range of 4--8 km~s$^{-1}$.
Note that the systemic velocity of source A is 3.1 km~s$^{-1}$.
The south part of the N-S ridge presents components in both velocity ranges, and its distribution coincides with that of the southeast part of the NW-SE outflow.
It is thus possible that the south part of the N-S ridge is in fact part of an outflow cavity rather than envelope structure.
Details and intensity integrated maps of both velocity ranges are given in Appendix~\ref{app:envout}.

The low spectral resolution of \ce{HNC} and \ce{HC3N}, as well as the clumpy nature of \ce{HC5N}, hinder a detailed kinematic analysis of the N-S ridge.
However, some insight can still be obtained.
A position velocity (PV) diagram is extracted from the \ce{HC5N} 35--34 datacube (channel width = 0.049 km~s$^{-1}$) with a position angle of 34$^{\circ}$ and a width of 7 pixels (1.12$\arcsec$), covering the extent of 33$\arcsec$ from the map edge to the east of source A (Fig.~\ref{fig:PVdiag}).
The resulting PV diagram, although clumpy, suggests a velocity increase as the distance to source A decreases (Fig.~\ref{fig:PVdiag}).
The increase in velocity inversely proportional to distance is characteristic of free-falling ($v \sim r^{-1/2}$) or infalling ($v \sim r^{-1}$) motion.
Such curves are plotted over the \ce{HC5N} PV diagram in Figure~\ref{fig:PVdiag} to show the trend, however, no estimate of the enclosed mass at any radius is made based on these curves.

Although our observations of \ce{HC3N} 10--9 and \ce{HNC} 1--0 were obtained with a low spectral resolution of only 0.8 km~s$^{-1}$, an indication of the velocity pattern can still be obtained by examining their centroid velocities along the N-S ridge.
Both molecules show velocity inversely proportional to distance from source A, however the velocities do not match with \ce{HC5N}.
The velocity discrepancy is due to several reasons.
From 33$\arcsec$ to 10$\arcsec$ away from source A, \ce{HNC} and \ce{HC3N} show inverse P Cygni spectral profiles (Fig.~\ref{fig:PVdiag}, Positions 1 and 2).
Although the locations are quite offset from source A, the continuum against which the line emission is absorbing is that of \iras.
As shown in \citet{bjerkeli2012}, inverse P Cygni profiles can be detected out to 100$\arcsec$ away from the continuum source.
The absorption becomes weaker with increased distance from the source, as is the case in the data presented here.
This inverse P Cygni profile is an indication of infalling material, but does not allow for centroid velocity analysis.
At offsets below 10$\arcsec$, the spectral profiles of \ce{HNC} and \ce{HC3N} become broad, while \ce{HC5N} shows several peaks (Fig.~\ref{fig:PVdiag}, Position 3). 
This is consistent with the complex outflow and envelope morphology, and thus a centroid velocity cannot be reliably determined.
Finally, about 5$\arcsec$ from source A, the spectral profile of \ce{HNC} shifts to P Cygni profile (Fig.~\ref{fig:PVdiag}, Position 4), indicative of outflowing material, while \ce{HC3N} presents a much broader spectral profile than \ce{HC5N}.
Hence, the kinematic analysis of \ce{HC3N} and \ce{HNC} is not inconsistent with that of \ce{HC5N}, but does not provide additional constraints.

The shift in velocity and spectral line profile along the north part of the N-S ridge, together with the inverse P Cygni profile and velocity inversely proportional to distance from source A, indicate that this is an infalling structure within the \iras~cloud core.
Further higher spectral resolution observations of \ce{HC3N} and \ce{HNC} with a larger field of view are necessary to provide a more detailed kinematic analysis.
In particular to disentangle the structure around source A, and be able to determine at what radius material is being delivered.
Higher spatial resolution is needed to further characterize the infall motion, and whether source A1, A2 or both are accreting the bulk of the material.

\subsection{Comparison with higher transitions}
\label{subsec:highJ}

\begin{figure*}
	\centering
	\includegraphics[width=0.9\linewidth]{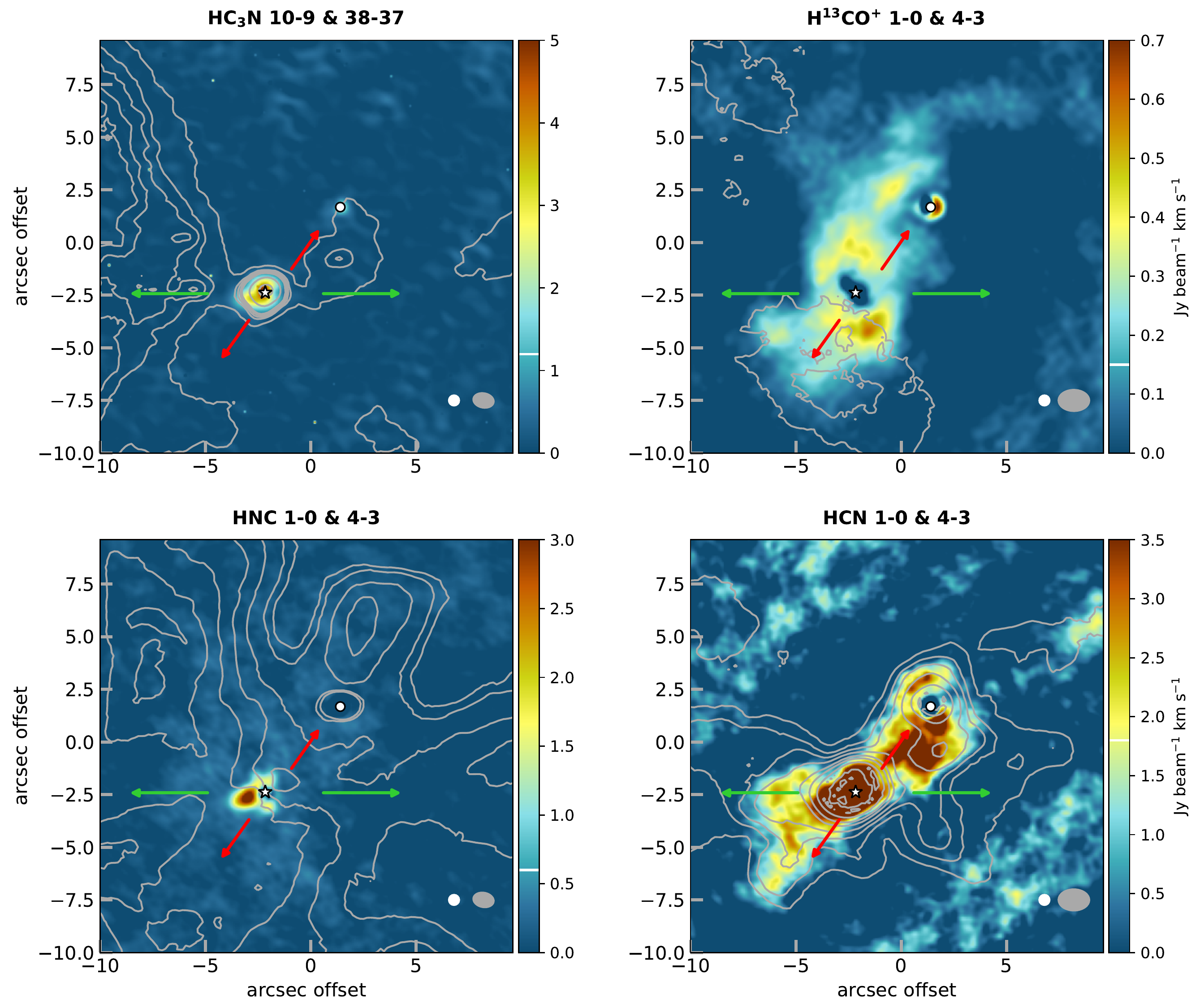}
	\caption{ALMA Band 7 data (PILS, colorscale) overlaid with Band 3 combined data (contours) for \ce{HC3N}, \ce{H^{13}CO+}, \ce{HNC}, and \ce{HCN}. The 3$\sigma$ level is indicated on the colorbar with a horizontal white or gray line. Panels are the size of the PILS Band 7 field of view. Contours are in steps of 3, 5, 7, 10, 15, 20, 30, 40, 50, and 80$\sigma$ where $\sigma$ = 70 mJy~beam$^{-1}$~km~s$^{-1}$ for \ce{HCN}; 3, 5, 10, 15, 20, 25, 30, 40, and 50$\sigma$ where $\sigma$ = 7 mJy~beam$^{-1}$~km~s$^{-1}$ for \ce{HNC}; 3, 5, 6, 7, 8, 15, 30, and 50$\sigma$ where $\sigma$ = 10 mJy~beam$^{-1}$~km~s$^{-1}$ for \ce{HC3N}; 2, 3, and 4$\sigma$ where $\sigma$ = 50mJy~beam$^{-1}$~km~s$^{-1}$ for \ce{H^{13}CO+}. The positions of \iras~A and B are marked with a star and circle, respectively. The arrows indicate the east-west (E-W, green) and northwest-southeast (NW-SE, red) outflow directions from source A. The white circle and gray ellipse show the Band 7 (PILS) and Band 3 beams, respectively.}
	\label{fig:moment0_B3B7}
\end{figure*}

Data from PILS include \ce{HNC} 4--3, \ce{HCN} 4--3, and three transitions of \ce{HC3N}: 37--36, 38--37, and 39--38 (Table~\ref{tab:lines}). 
No line emission is detected from the transitions of \ce{HC5N} within the frequency range of the PILS program (329.147 -- 362.896 GHz).
The PILS data, which combined 12-m array and ACA observations, recover scales from 0.5" up to 13" with a field of view of 16". 
Thus only the inner 16" of the Band 3 observations can be compared.
Figure~\ref{fig:moment0_B3B7} presents overlaid Band 7 (color) and Band 3 (contours) intensity integrated maps.

The three Band 7 transitions of \ce{HC3N} all show an emission peak on source A, but no further extended structures are detected.
The brightest transition, \ce{HC3N} 38--37, is included in Figure~\ref{fig:moment0_B3B7} for comparison.
Similarly, \ce{HNC} 4--3 peaks about 1$\arcsec$ east of source A, but presents no other significant extended emission.

In the direction of the southeast outflow lobe, \ce{H^{13}CO+} presents emission in both 1--0 and 4--3 transitions (Fig.~\ref{fig:moment0_B3B7}) in a velocity range between 4 to 7 km~s$^{-1}$ (Fig.~\ref{fig:moment0_outenv}).
The velocity range and spatial agreement of both transitions suggests the southeast \ce{H^{13}CO+} arises from the outflow cavity walls.
North of source A, \ce{H^{13}CO+} 4--3 traces the material between sources A and B, along the dust bridge, but not along the region where the N-S ridge is detected like the 1--0 transition.
Despite the different spatial distributions, both transitions are present between 1 and 4 km~s$^{-1}$ (Fig.~\ref{fig:moment0_outenv}), indicating their envelope nature.
The inner circular region around source A where \ce{H^{13}CO+} is not present (Fig.~\ref{fig:moment0_B3B7}) coincides with the 100 K radius \citep{jacobsen2018}, tracing the water snowline (e.g., \citealt{vanthoff2018,leemker2021}). 
It is worth highlighting that the east-west outflow does not present \ce{H^{13}CO+} emission in either transition.
\ce{HCN} 4--3 traces material extending between sources A and B.
While the \ce{HCN} 1--0 emission traces both outflows driven by source A, the 4--3 transition mainly arises from the northwest-southeast outflow and part of the west outflow lobe cavity wall \citep{vanderWiel2019}.

\subsection{Line ratios}
\label{subsec:ratios}

\begin{figure*}
	\centering
	\includegraphics[width=0.92\linewidth]{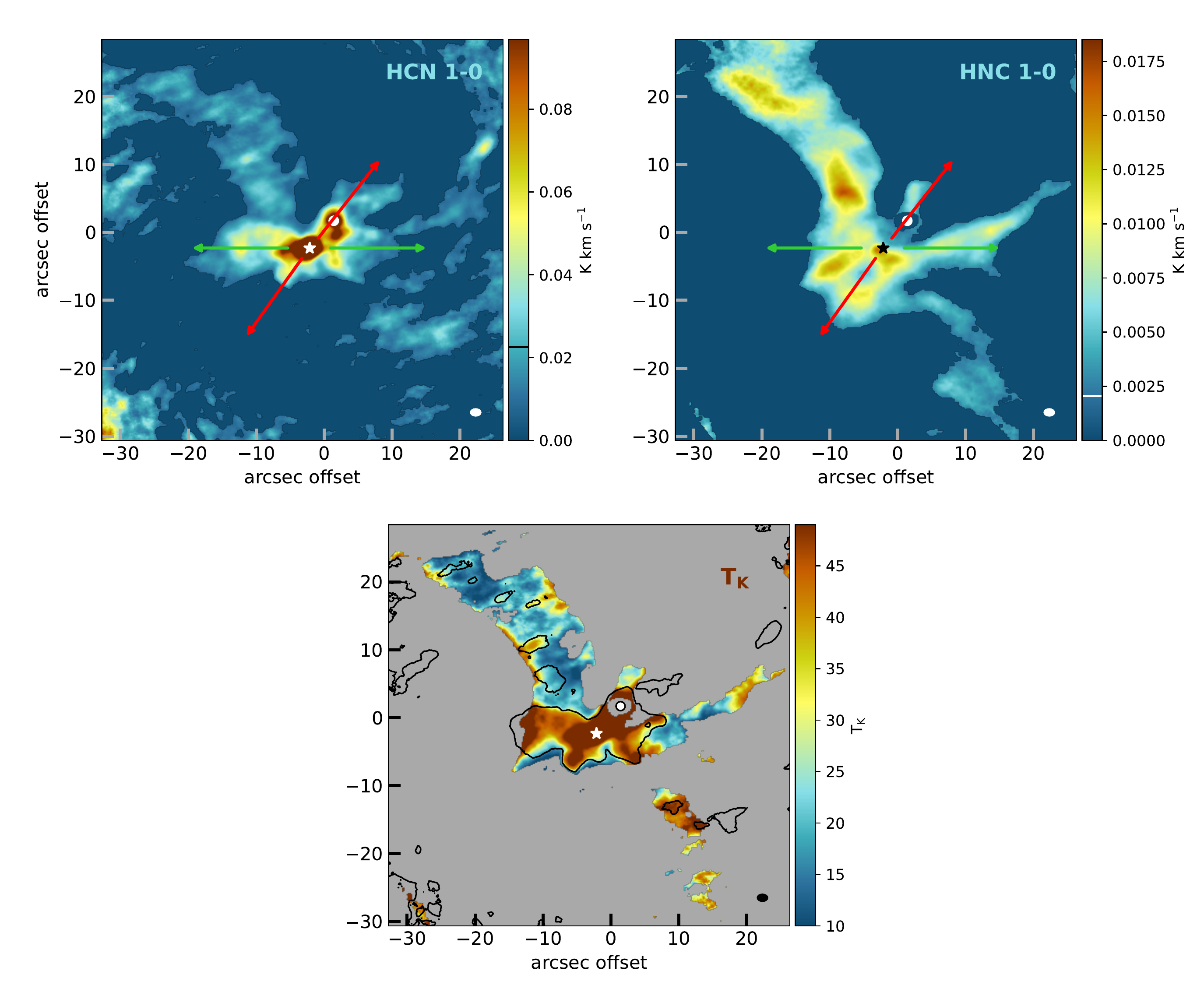} 
	\caption{Intensity integrated map of \ce{HCN} (\textit{top left}) and \ce{HNC} (\textit{top right}) in K~km~s$^{-1}$. The \ce{HCN} map was masked at 1$\sigma$ (0.07 Jy~beam~km~s$^{-1}$), while \ce{HNC} was masked at 3$\sigma$ (0.02 Jy~beam~km~s$^{-1}$), then converted to brightness temperature. The 3$\sigma$ level is indicated on the colorbar with a horizontal black or white line. Source A and B positions are marked with star and circle, respectively. The arrows indicate the east-west (E-W, green) and northwest-southeast (NW-SE, red) outflow directions from source A. The white ellipse on the bottom left indicates the map beamsize. The gas kinetic temperature $T_{\rm K}$ map, obtained from \ce{HCN}/\ce{HNC} and Eq.~\ref{eq:leq4} and \ref{eq:bt4}, is shown in the bottom panel. The colorscale is limited by the \ce{HNC} 3$\sigma$ level. The black contours indicate the \ce{HCN} 3$\sigma$ level, outside of which $T_{\rm K}$ is an upper limit. The lack of HCN emission indicates that the N-S ridge is mainly composed of cold gas, which is constrained by the ratio upper limits. Temperatures above 40 K are not well constrained by the \ce{HCN}/\ce{HNC} ratio, hence the temperature scale only runs up to 49 K.}
	\label{fig:ratio}
\end{figure*}

Spherical envelope models of \iras~indicate gas and dust temperatures of $\sim$40, 30 and 20 K at radii of 500, 1000 and 2000 AU, respectively \citep{schoier2002,crimier2010}. 
Three dimensional modeling of the dust emission (including the spectral energy distribution, SED) in the \iras~envelope \citep{jacobsen2018} shows dust temperatures of 30 K out to a radius of 750 AU to the east, south and west of source A.
While dust temperatures drop below 30 K beyond 500 AU north of source A in the same model.
Gas temperatures derived from molecular line emission observations within 1000 AU of source A \citep{murillo2018} range from 20 K (\ce{DCO+} at the disk edge) to 150 K (outflow cavity traced in \ce{c-C3H2}).
The physical conditions of the low-$J$ molecular line emission presented in this work are determined in two ways: 1) \ce{HCN}/\ce{HNC} abundance ratio, and 2) line ratio of the same molecule in two transitions.

Recently, \citet{hacar2020} demonstrated empirically that the \ce{HCN}/\ce{HNC} ratio, both 1--0 transitions, is a probe of gas temperature in the range of 14 to 40 K.
The \ce{HCN}/\ce{HNC} line intensity ratio is related to the gas kinetic temperature with the following relations derived by \citet{hacar2020}:
\begin{equation}
	\label{eq:leq4}
	T_{\rm K} = 10 \times \frac{I(\ce{HCN})}{I(\ce{HNC})} \hspace{22.5mm} if \hspace{3mm} 1 < \frac{I(\ce{HCN})}{I(\ce{HNC})} \leq 4
\end{equation}
\begin{equation}
	\label{eq:bt4}	
	T_{\rm K} = 3 \times \left( \frac{I(\ce{HCN})}{I(\ce{HNC})} -4 \right) + 40 \hspace{15mm} if \hspace{3mm} \frac{I(\ce{HCN})}{I(\ce{HNC})} > 4.
\end{equation}
Both \ce{HCN} and \ce{HNC} are formed in a 1:1 ratio from \ce{HCNH+} dissociative recombination. The destruction of \ce{HNC} in reactions with \ce{H} and \ce{O} is more efficient with temperature. At higher temperatures, \ce{HNC} is isomerized, thus increasing the abundance of \ce{HCN}.

To produce a ratio map of \ce{HCN} and \ce{HNC} for \iras, the images need to have the same range of recovered scales.
An \ce{HNC} map from our combined Band 3 data was generated using the same $(u,v)$ range as the \ce{HCN} data.
The resulting \ce{HNC} map was then convolved with the same beam as the \ce{HCN} map, 1.34$\arcsec\times$0.94$\arcsec$.
Intensity integrated maps of \ce{HCN} and \ce{HNC} were then generated, clipped at the 3$\sigma$ level for \ce{HNC} and 1$\sigma$ level for \ce{HCN}. The maps were then converted to units of K~km~s$^{-1}$ (Fig.~\ref{fig:ratio}).
The mean flux density in mJy~beam$^{-1}$ km~s$^{-1}$ is converted to brightness temperature in K km~s$^{-1}$ by 
$T_{\rm mb}$ = 1.36~$\lambda^{2}/\theta^{2}~I_{\rm \nu}$,
where $\lambda$ is the wavelength in centimeters of the molecular transition, $\theta$ is the synthesized beam in arcseconds, and $I_{\rm \nu}$ is the flux density in mJy~beam$^{-1}$ km~s$^{-1}$.
The \ce{HCN}/\ce{HNC} ratio map is shown in Figure~\ref{fig:ratio}, where the colorscale is limited by the \ce{HNC} 3$\sigma$ level, and the \ce{HCN} 3$\sigma$ level is indicated by black contours.
Temperatures derived from ratios outside of the black contours are upper limits.
The lack of \ce{HCN} emission indicates that the N-S ridge is mainly composed of cold gas (T $\sim$ 20 K). This is consistent with the chemistry of \ce{HCN} and \ce{HNC}, and reflected by the upper limits of the \ce{HCN}/\ce{HNC} ratio.


Along the N-S ridge, the \ce{HCN}/\ce{HNC} ratio indicates a mean gas temperature of $\sim$25 K when considering only points with \ce{HCN} above 3$\sigma$, and $\sim$20 K when considering all points along the N-S ridge shown in Figure~\ref{fig:ratio}. 
For regions where both isomers are present, the typical mean error is 9 K.
There are a few pixels with \ce{HCN}/\ce{HNC} ratios between 4 and 8.5, along the edges of the N-S ridge and \ce{HCN} blobs, which would suggest pockets of gas $\geq$40 K.
Not considering these pixels with higher ratios, the gas temperature ranges between 10 -- 30 K along the N-S ridge.
One caveat that should be considered is that the \ce{HCN} blobs are evenly separated and aligned from northeast to southwest, which make them prone to be artifacts of the uv-coverage, rather than real structure.
This suggests that the N-S ridge is mainly composed of cold gas, with perhaps a few clumps of warmer gas.
The \ce{HCN}/\ce{HNC} ratio along the west ridge gives an upper limit mean of $\sim$30 K (Fig.~\ref{fig:ratio}).
The origin of the warm pockets along the N-S ridge is not clear, given the lack of a heating source or dust, and the low envelope dust and gas temperatures derived from models (but see Sect.~\ref{subsec:scene}).

In the region around sources A and B, where \ce{HCN} emission is predominant, the \ce{HCN}/\ce{HNC} ratio ranges between 1.5 and 82, implying temperatures from 15 K to well above 40 K.
However, kinetic temperatures above 40 K are not well constrained by the empirical method of \citet{hacar2020}. 
This temperature structure is consistent with the envelope dust and gas temperature derived from models (e.g., \citealt{crimier2010,jacobsen2018}), and that the bulk of protostellar heating escapes through the outflow cavities.


Constraints on the density of the N-S ridge, as well as the inner region around sources A and B can be obtained by comparing the low-$J$ (Band 3) and higher-$J$ (Band 7) transitions of \ce{HNC}, \ce{HC3N} and \ce{HCN}, together with the gas kinetic temperatures derived from the \ce{HCN}/\ce{HNC} ratio.
Details of the analysis are presented in Appendix~\ref{app:ratios} and Figure~\ref{fig:highvslowratios}.
The comparison indicates that the N-S ridge has \ce{H2} densities of $\sim$10$^{5}$ cm$^{-3}$, while the inner region has one to two orders of magnitude higher \ce{H2} densities for temperatures between 10 and 40 K.
These physical parameters are consistent with envelope models of \iras.

\section{Discussion}
\label{sec:discussion}

The observations presented in this work reveal two extended structures present in the cloud core of \iras, both related to source A and traced in several molecular species.
The physical and chemical conditions of the two structures, the N-S and west ridges, are summarized in this section and placed in context of previous work.
The main observational result of this work is the uneven distribution of material within the cloud core of a multiple protostellar system

\subsection{Accretion stream}
\label{subsec:stream}

The N-S ridge is composed of cold gas ($\sim$15 K), with a few clumps at $\leq$40 K, and presents a layered molecular gas structure, but no dust continuum.
The N-S ridge is best traced in \ce{HNC}, \ce{HC3N}, and \ce{HC5N}.
Partial \ce{N2H+} and faint \ce{H^{13}CO+} along the N-S ridge, together with \ce{CO} only tracing the outflows driven by source A \citep{vanderWiel2019}, indicate that these molecules are not good tracers of accretion flows for \iras.
The N-S ridge is only present at envelope velocities (0.8 -- 4.3 km~s$^{-1}$).
Kinematic analysis of \ce{HC5N}, \ce{HC3N} and \ce{HNC} suggest infalling gas along the north part of the N-S ridge, in particular the presence of \ce{HNC} and \ce{HC3N} inverse P Cygni profiles.
Based on these results we propose that the northern part of the N-S ridge is a cold accretion structure transferring material from the outer parts of the cloud core onto \iras~A.
The nature of the southern part of the N-S ridge cannot be fully determined with the current data due to spatial overlap with the southeast lobe of the NW-SE outflow.
We need higher spectral resolution observations of \ce{HC3N} and \ce{HNC} to confirm the kinematic structure of the N-S ridge.
Interestingly, no connection between the N-S ridge and source B is detected.

Accretion flows, or accretion streamers, have been previously detected in other embedded sources at a 10$^{3}$ -- 10$^{4}$ AU scales \citep{alves2019,pineda2020,alves2020}.
In addition, models of multiple protostellar formation also show such long and narrow structures readily forming in the cloud core of these systems (e.g., \citealt{offner2010,kuffmeier2019}).
Thus, the scenario of the N-S ridge being an accretion streamer feeding material onto \iras~A is likely.
However, the N-S ridge is only transferring material to one of the components in the wide multiple protostellar system.
Continuous delivery of envelope gas onto the circumstellar environment of source A, but not source B, could explain the difference in multiplicity and bolometric luminosity (18 L$_{\odot}$ vs 3 L$_{\odot}$, \citealt{jacobsen2018}) between the two sources.
This is supported by a rough estimate of accretion rates, the derivation of which is detailed in Appendix~\ref{app:accretion}.
The north part of the N-S ridge has an estimated accretion rate of  7$\times$10$^{-7}$ to a few 10$^{-6}$ M$_{\odot}$~yr$^{-1}$.
Conversely, assuming that accretion luminosity is equal to the bolometric luminosity of source A, $L_{acc}$ = $L_{bol}$ = 18 L$_{\odot}$, the accretion rate is about 8$\times$10$^{-7}$ M$_{\odot}$~yr$^{-1}$.
With the same luminosity assumption, source B ($L_{bol}$ =  3 L$_{\odot}$) would have an accretion rate of a few 10$^{-7}$ M$_{\odot}$~yr$^{-1}$.
Preferential accretion of material could be key in multiple protostellar system formation and evolution \citep{murillo2018c}. 
In particular, uneven delivery of material in wide multiple protostars (separations larger than disk radius) could provide insight into systems with components at different evolutionary stages \citep{murillo2016}.

It should be stressed that the N-S ridge is cleanly traced in \ce{HC3N}, suggesting it could be a good tracer for cold gas accretion flows in the cloud core of embedded protostars.
Indeed, \ce{HC3N} traces the accretion flow in Per-emb 2 (IRAS 03292+3039, \citealt{pineda2020}).
However, the N-S ridge is also well traced in \ce{HNC}, and \ce{HC5N}, partially in \ce{N2H+}, with faint \ce{H^{13}CO+} and \ce{HCN} emission.
Tracing accretion flows in cloud cores using several of the same molecules used to trace filamentary structures on molecular cloud scales (R $>$ 10$^{4}$ AU; \citealt{arzoumanian2013,hacar2013,fernandez2014,hacar2017,hacar2020}) makes it possible to eventually trace the transfer of material from parsec scales down to protostellar scales.
In addition, \ce{HNC} is commonly detected in embedded protostellar sources ($\sim$7000 AU; \citealt{murillo2018c}), and thus also a potential good tracer for cold gas accretion flows in cloud cores.
This is important, since not all embedded protostars may have carbon chains present in the cloud core (e.g., \citealt{murillo2018c}), in particular cyanopolyynes, leading to non-detections of accretion flows if only this tracer were used.
In general, the physical conditions of a cloud core should be considered when determining which molecular species to use for tracing accretion flows.


\subsection{Outflow cavity walls}
\label{subsec:cavitywalls}

The west ridge, detected in \ce{HC3N}, \ce{HC5N}, and \ce{HNC}, is related to the E-W outflow, and composed of cold gas.
The spatial and velocity distributions of the west ridge are consistent with the red-shifted, west outflow cone detected in \ce{^{12}CO} 3--2 (8 to 14 km~s$^{-1}$), and the northern cavity wall of the same lobe detected in other molecules (4 to 7 km~s$^{-1}$) \citep{vanderWiel2019}.

The molecules \ce{N2H+} and \ce{H^{13}CO+} show emission along the southeast lobe of the NW-SE outflow.
The chemistry of these species provide gas temperature constraints.
Given that \ce{N2H+} is present when \ce{CO} is frozen out (gas and dust temperature below 20--30 K), the emission traces the outer cavity wall rather than the outflow itself.
On the other hand, \ce{H^{13}CO+} requires \ce{CO} in the gas phase (temperatures above 20--30 K) to form, and reacts with water vapor (gas temperature above 100 K).
Thus, \ce{H^{13}CO+} traces a warmer part of the outflow cavity with a temperature range between 30 and 100 K, which is consistent with the gas temperature range inferred from \ce{c-C3H2}, 120 -- 155 K \citep{murillo2018}.


Comparing the observed molecular species and derived conditions indicates that the NW-SE outflow is significantly warmer than the E-W outflow.
This is consistent with the E-W outflow being more prominent in the Band 3 observations, than in the Band 7 observations.
The contrast of molecular species tracing each outflow, and respective cavity wall, demonstrate the layered nature of these objects, and the impact of the driving source on the chemical structure of the cavity.
This could provide a further constraint in determining which source in the close binary A1+A2 drives each outflow.

\begin{figure*}
	\centering
	\includegraphics[width=0.93\linewidth]{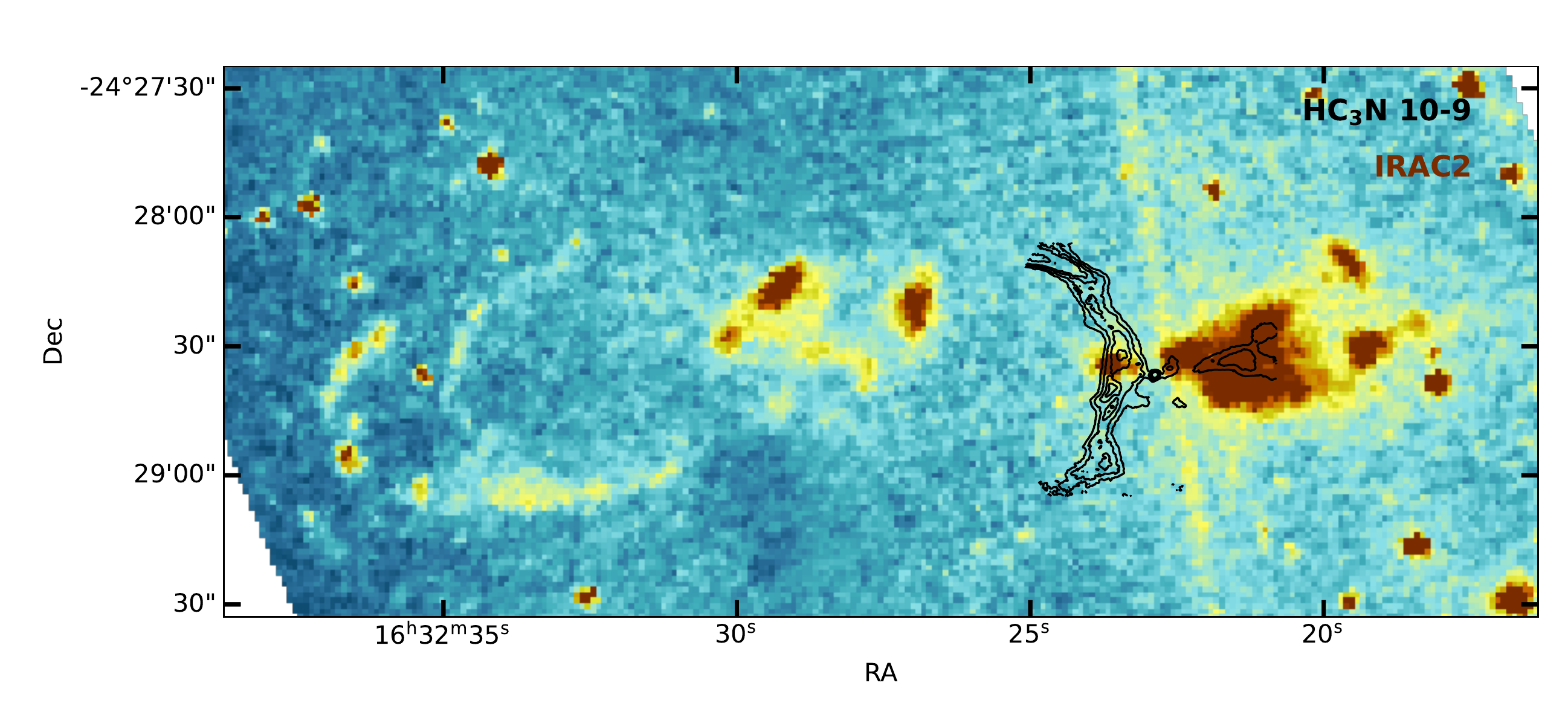}
	\caption{\textit{Spitzer} IRAC2 image of \IRAS~(colorscale), overlaid with \ce{HC3N} (contours). The IRAC2 image shows the E-W outflow cavities, but does not show structures relating to the NW-SE outflow. The \ce{HC3N} west ridge matches with the west lobe of the E-W outflow. The N-S ridge is not related to the outflow cavity walls, but its chemistry may be impacted by the east lobe of the E-W outflow.}
	\label{fig:spitzer}
\end{figure*}

\subsection{Large scale scenario}
\label{subsec:scene}

The layered molecular gas structure of the N-S ridge and the presence of \ce{HC3N} and \ce{HC5N} at these scales suggest that the N-S ridge gas is externally heated. 
A possible source of heating can be the UV-emitting jet shocks within the east cavity of the E-W outflow seen in large scale \textit{Spitzer} IRAC2 images (Fig.~\ref{fig:spitzer}).
Heating of the N-S ridge from one side also explains the layered nature of the gas, with \ce{N2H+} located furthest from the outflow cavity, consequently implying a temperature gradient from east to west.
It could be argued that the N-S ridge is a limb brightening around the outflow cavity. 
However, the physical conditions and chemistry along the N-S ridge suggest otherwise.
The velocity range of the N-S ridge already indicates that it is not part of the outflow cavity, but the envelope.
In addition, the different chemical structures of the N-S ridge and the west ridge, which is part of the E-W outflow cavity wall, also provide evidence against the limb brightening scenario of the N-S ridge.
The limited kinematic analysis possible with the present data hint at the infalling nature of the N-S ridge, but higher spectral resolution is needed to probe the kinematics of the ridge.

Something worth noting is the lack of carbon chain molecules (\ce{C_{n}H_{n-1}}), while cyanopolyynes (\ce{HC_{n}N} where n = 3,5,7...) are readily present.
A transition of \ce{c-C3H2} at 93.16 GHz (log$_{10}$ $A_{ij}$ = -4.76, $E_{up}$ = 92 K) falls in the frequency range of our Band 3 observations, but it is not detected.
Below, the possible reasons for the lack of extended cyclopropenylidene emission are briefly explored.

The critical density of \ce{c-C3H2} at 93.16 GHz lies between 10$^3$ and 10$^4$ cm$^{-3}$ for a temperature range of 10 to 100 K, while that of \ce{HC3N} 10--9 is about 10$^4$ cm$^{-3}$ for a similar temperature range.
Hence, the lack of \ce{c-C3H2} is not due to density or gas temperature.
In Band 7 observations, \ce{c-C3H2} is detected mainly within the NW-SE outflow cavity with gas temperatures above 100 K and spatially anti-correlated with \ce{C2H} \citep{murillo2018}.
Hence, the lack of detected \ce{c-C3H2} may be due to chemistry, rather than physical conditions.
It is interesting to note that \ce{HC3N} is readily formed through the following gas phase reactions \citep{loison2014,loison2017}:
\begin{equation*}
	\cee{C2H + HNC -> H + HC3N}
\end{equation*}
\begin{equation*}
	\cee{N + c-C3H2 -> H + HC3N}
\end{equation*}
with the former and latter reactions also being valid for \ce{HCN} and \ce{l-C3H2}, respectively.
The latter reaction, while minor, is non-negligible \citep{loison2017}, and the abundance of nitrogen in the gas phase as evidenced by \ce{N2H+}, makes it even more viable.
Further interaction between \ce{C2H} and \ce{HC3N} can lead to the formation of \ce{HC5N}.
Similarly, the Band 3 observations presented here show that \ce{HNC} is also present.
Hence, the lack of carbon chains and presence of cyanopolyynes are likely due to the above reactions.
This then explains the layered structure observed along the N-S ridge, in particular the offset between \ce{HNC} and the cyanopolyynes.
Furthermore, it would also provide a reason why \ce{HC3N} and \ce{HC5N} are present in the E-W outflow, but are much weaker in the NW-SE outflow which does present \ce{c-C3H2} emission in Band 7 observations.
Given the presence of \ce{HNC} and \ce{HCN} in the southeast outflow cavity would suggest that the non-detection of \ce{C2H} \citep{murillo2018} is due to the reaction between these molecules to produce \ce{HC3N}.

The results presented in this work demonstrate the importance of observing a broad range of frequencies and molecular lines toward protostellar systems, and the capacity of ALMA to probe cloud core structure, in particular with short baselines.
To identify accretion flows or streams, and differentiate them from outflow structures, the cloud core needs to be characterized at a range of scales with molecular species tracing cold and warm gas. 
Sufficient spectral resolution is also crucial to examine the velocity structure.

\section{Conclusions}
\label{sec:conclusions}
We present ALMA Band 3 (3 mm) observations of \IRAS.
The data presented in this work include observations of the low-$J$ transitions of cyanopolyynes \ce{HC3N}, \ce{HC5N}, along with \ce{HNC}, \ce{HCN}, \ce{H^{13}CO+}, \ce{N2H+}, and \ce{SiO}.
Further comparison with higher-$J$ transitions was done using ALMA Band 7 data from the PILS program \citep{jorgensen2016}.

The results of this work can be summarized in the following points:
\begin{enumerate}
	\item Extended, cold gas ($\leq$20 K) structures from scales of a few 100s AU up to 6000 AU (40$\arcsec$) are detected with our Band 3 observations toward \iras~with a spatial resolution of 130 AU.
	Two main structures are identified: a N-S ridge, and a west ridge. 
	These are traced best in \ce{HC3N}, \ce{HNC}, and \ce{HC5N}. 
	The molecules \ce{H^{13}CO+} and \ce{N2H+} only partially trace the N-S ridge.
	No carbon chain molecules are detected.
	These structures do not present a counterpart in dust continuum, or in higher transitions (Band 7) of the same molecular species. 
	\item Our results dwell into the detailed structure of sources A and B, providing an updated scenario of this system.
	We propose the following scenario based on the results of the current data.
	Envelope material from a few 1000 AU is being delivered, and accreted, by \iras~A, while \iras~B can accrete only its immediate circumstellar material.
	This scenario could explain the different bolometric luminosities of sources A and B (18 and 3 L$_{\odot}$, respectively), and the binarity of \iras~A.
	\item The molecules tracing the N-S ridge are slightly spatially offset relative to each other, generating a layered structure. 
	The layering and lack of carbon chains along the N-S ridge can be explained by chemical processes, namely the reaction of carbon chains with \ce{HNC} and nitrogen to produce \ce{HC3N} and \ce{HC5N}.
	Furthermore, the chemical structure of the N-S ridge helps in determining its nature.
	\item The west ridge traces part of the northern cavity wall of the west outflow lobe. 
	The difference in molecular species present in each of the outflows driven by \iras~A, and their cavity walls, suggests different excitation conditions produced by the driving sources.
\end{enumerate}

The results presented here show the importance of streamers in shaping individual protostellar components, and their circumstellar environment, in multiple protostellar systems.
This work also demonstrates the importance of multi-frequency observations and a range of spatial resolutions for a complete picture of the physical structure of protostellar systems, multiple or single.
Future observations of the extended protostellar cloud core environment comparing low- and high-$J$ lines will prove invaluable to further understand how material is structured and distributed among components in multiple protostellar systems.
Combining this information from cloud core to disk scales will enable a better understanding of the formation and evolution of multiple protostellar systems.

\begin{acknowledgements}
	This paper made use of the following ALMA data: ADS/JAO.ALMA 2013.1.00278.S, 2015.1.01193.S, and 2017.1.00518.S. ALMA is a partnership of ESO (representing its member states), NSF (USA), and NINS (Japan), together with NRC (Canada) and NSC and ASIAA (Taiwan), in cooperation with the Republic of Chile. The Joint ALMA Observatory is operated by ESO, AUI/NRAO, and NAOJ. 
	This project has received funding from the European Research Council (ERC) under the European Union’s Horizon 2020 research and innovation programme (grant agreement No. 851435 to A.H., and No. 646908 to J.K.J.). The research of J.K.J. is furthermore supported by the Independent Research Fund Denmark (grant number 0135-00123B).  
\end{acknowledgements}

\bibliographystyle{aa}
\bibliography{extended_structure.bib}

\begin{appendix}

\section{Spectra}
\label{app:spectra}

\begin{figure*}
	\centering
	\includegraphics[width=\textwidth]{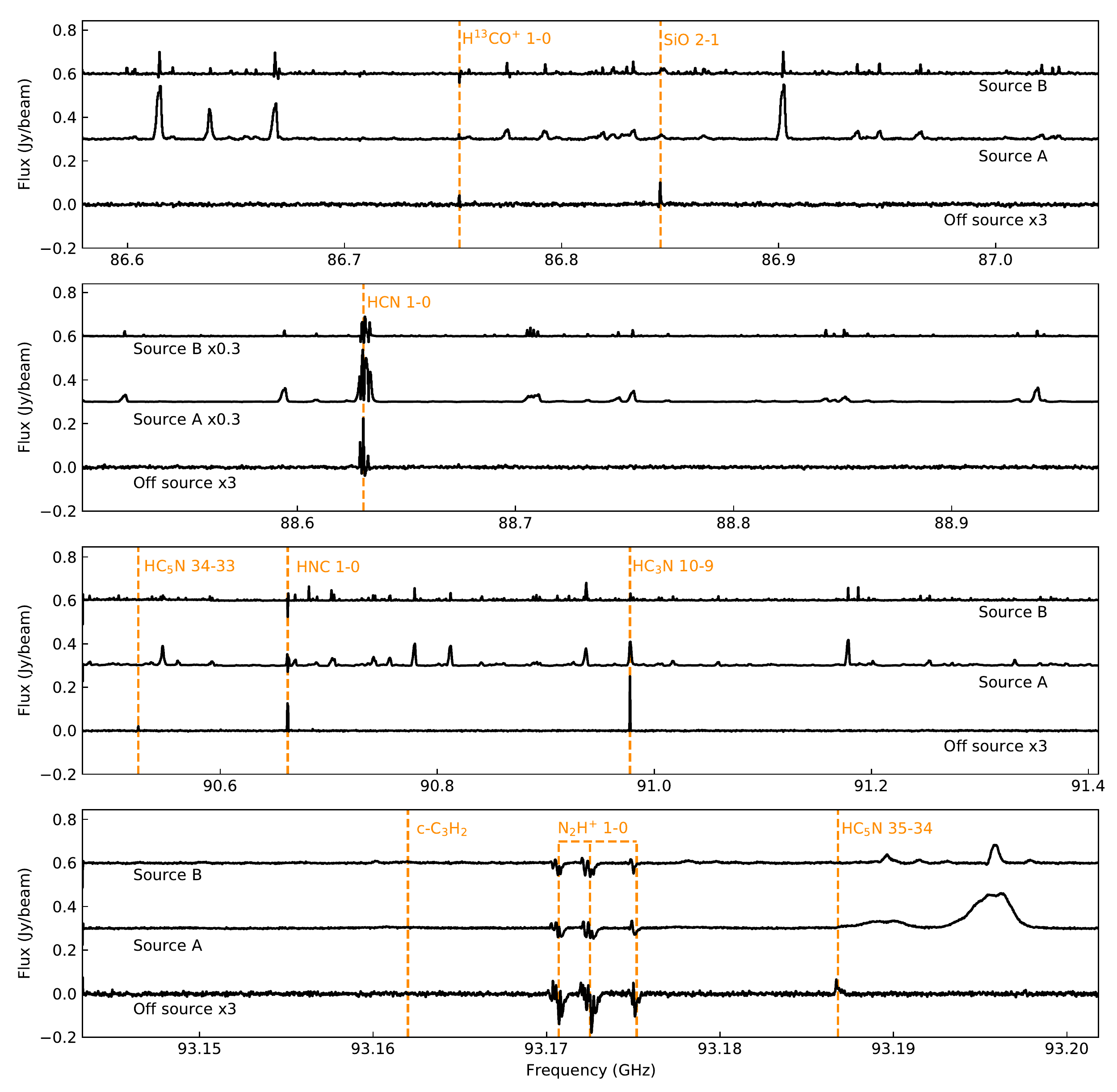}
	\caption{ALMA Band 3 spectra extracted from three positions: source A, source B ($\sim$1$\arcsec$ off), where the peak of emission is located, and an off-source position 12$\arcsec$ to the east of source B. Positions are shown in the right panel of Fig.~\ref{fig:continuum}. For the purpose of comparison, the off-source spectra is multiplied by a factor of 3, while the \ce{HCN} spectra for source A and B are multiplied by 0.3. The molecular species which show up in the off-source position are labeled. In addition, \ce{c-C3H2} is also labeled for reference. Note the chemical richness of the source positions in contrast to the off-source position.}
	\label{fig:spectra}
\end{figure*}

ALMA Band 3 spectra were extracted from three locations shown in Figure~\ref{fig:continuum}: on source A, $\sim$1$\arcsec$ west from source B, and an off-source position 12$\arcsec$ east of source B and away from the outflow directions.
Spectra was extracted using a single pixel of the image, with the pixel size being 0.16$\arcsec$.
The resulting spectra are presented in Figure~\ref{fig:spectra}.
The extracted spectra show both source A and B to be line rich, with line densities of $\sim$120 lines/GHz at 86, 87 and 91 GHz, and 34 lines/GHz at 93 GHz.
In contrast, the off-source position has a line density of 8 lines/GHz.

\section{Kinematics: envelope vs. outflow}
\label{app:envout}

\begin{figure*}
	\centering
	\includegraphics[width=0.75\textwidth]{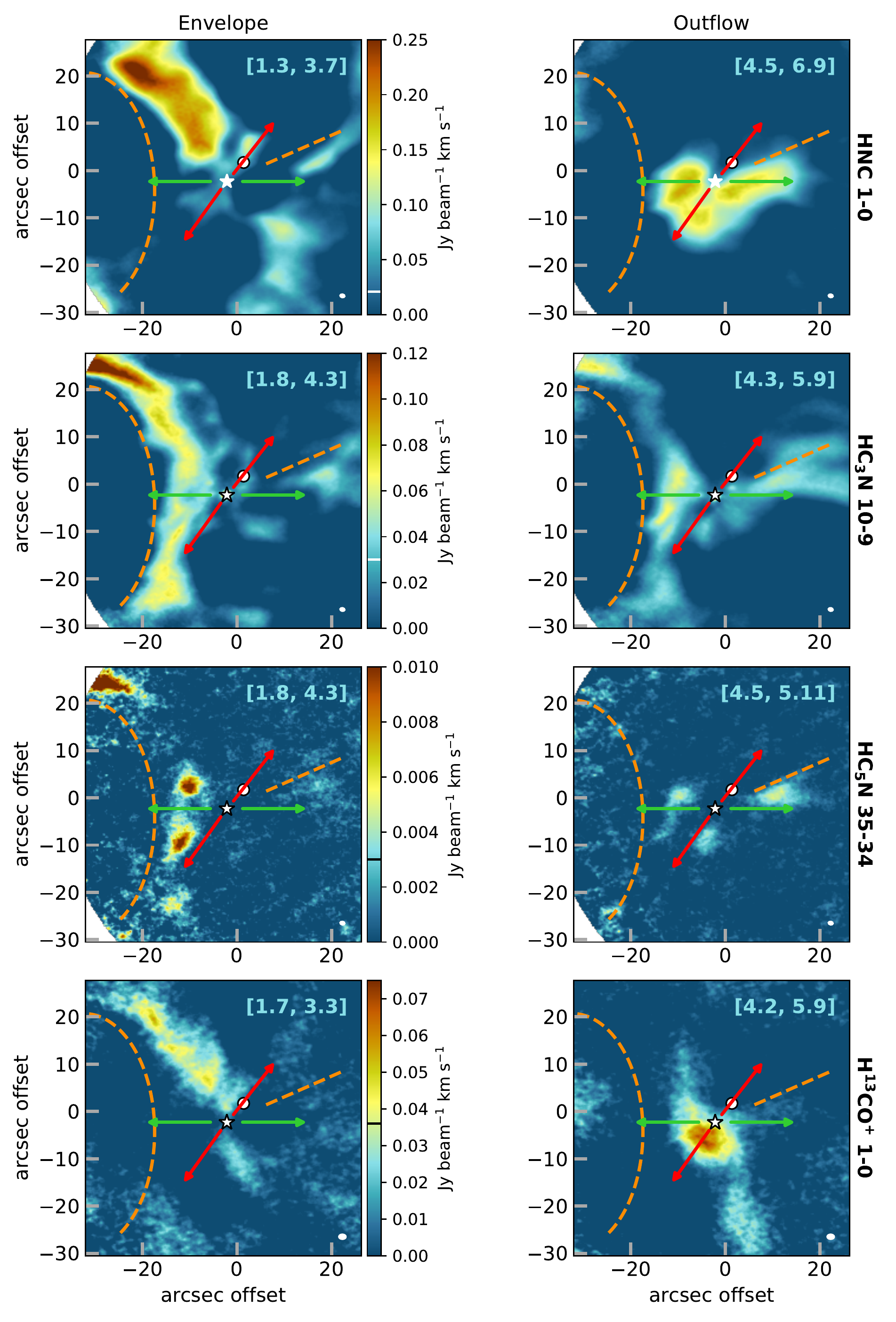}
	\caption{Intensity integrated maps aiming to disentangle the envelope and outflow contributions to the observed emission. The velocity range over which each map was integrated is given in square brackets in units of km~s$^{-1}$. The positions of \iras~A and B are marked with a star and circle, respectively. The arrows indicate the east-west (E-W, green) and northwest-southeast (NW-SE, red) outflow directions from source A. The white ellipse on the bottom right shows the beam. }
	\label{fig:moment0_outenv}
\end{figure*}

The N-S and west ridges present different velocity ranges (Fig.~\ref{fig:moment0_outenv}).
The N-S ridge is present only at velocities between 1.3 and 4.3 km~s$^{-1}$.
Given the systemic velocities of source A and B, 3.1 and 2.7 km~s$^{-1}$ respectively, the N-S ridge mainly traces envelope material, particularly in the north part of the N-S ridge.
In contrast, the west ridge is characterized by emission at the velocity range of 4--8 km~s$^{-1}$, which is consistent with the observed velocity range for the red part of the outflows \citep{vanderWiel2019}.
Detailed kinematic analysis of both ridges cannot be performed with \ce{HC3N}, \ce{HNC}, or \ce{H^{13}CO+} due to the low spectral resolution of the data, 0.8 km~s$^{-1}$.
While \ce{N2H+} has a higher spectral resolution of 0.049 km~s$^{-1}$, strong absorption due to filtering prevent us from extracting the detailed kinematic information from the cold gas traced by \ce{N2H+}.

\section{Ratios of high-$J$ vs low-$J$ transitions}
\label{app:ratios}

The non-detection of Band 7 high-$J$ transitions of \ce{HNC} and \ce{HCN} along the N-S ridge provides upper limits on the physical conditions of this structure.
While the line ratio of \ce{HC3N} would provide further constraints, the available molecular data file only covers up to J=37 \citep{faure2016}. 
Of the Band 7 \ce{HC3N} transitions, the 38--37 shows the most extended emission, while the 37--36 transition is concentrated on source A.
Comparison of the low- and high-$J$ transitions is made within the Band 7 field of view around source A and where the N-S ridge is located (Fig.~\ref{fig:moment0_B3B7}).
If the noise level of the Band 7 observations is taken as the upper limit for the high-$J$ \ce{HCN}, \ce{HNC} and \ce{HC3N} transitions, the peak intensity in Band 7 is at least an order of magnitude lower than in Band 3.

\begin{figure*}
	\centering
	\includegraphics[width=0.95\textwidth]{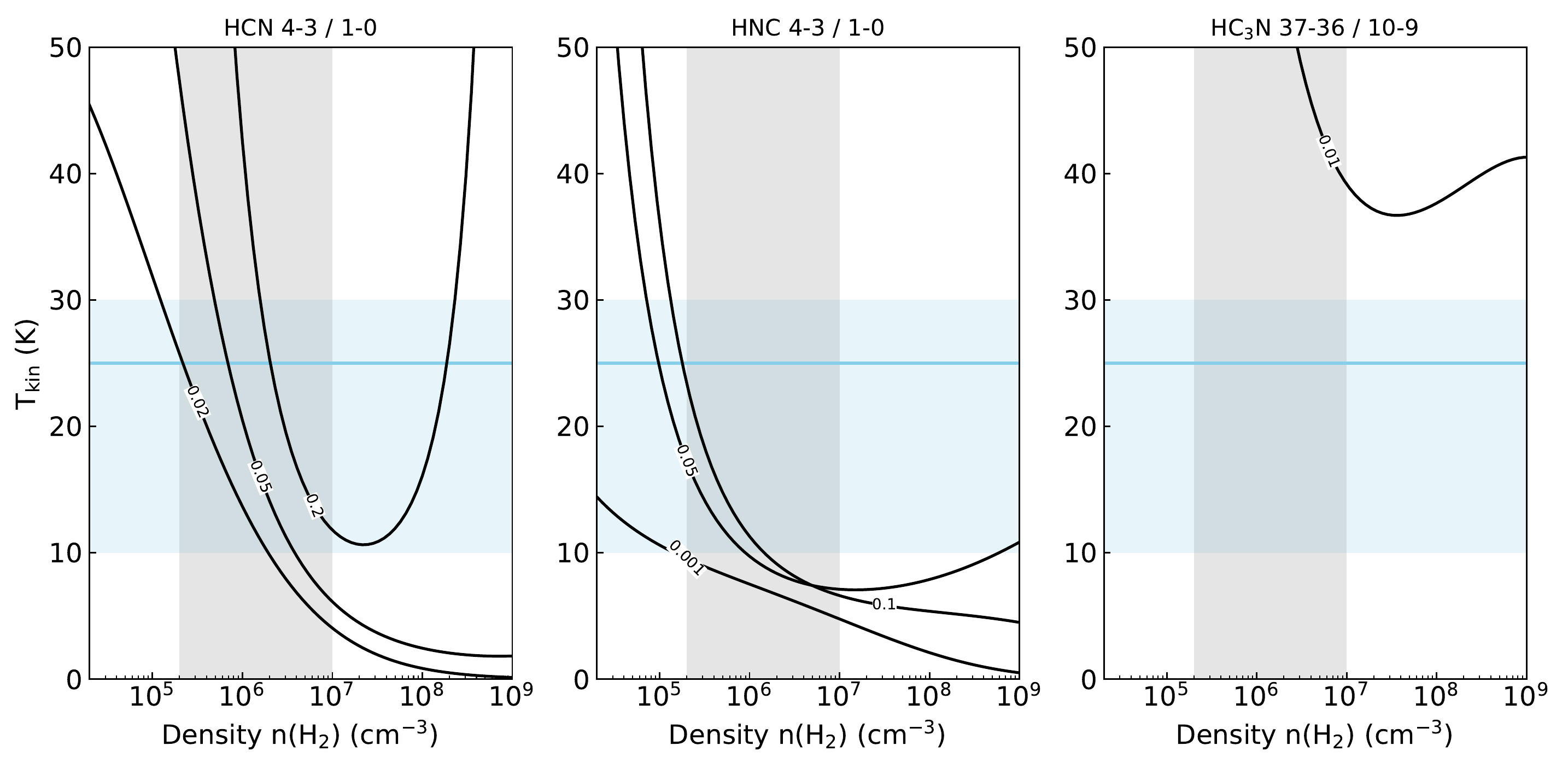}
	\caption{Calculated brightness temperature ratios for \ce{HCN} 4--3/1--0 (\textit{left}), \ce{HNC} 4--3/1--0 (\textit{center}), and \ce{HC3N} 37--36/10--9 (\textit{right}). Contours show modeled ratios assuming a column density of 10$^{13}$ cm$^{-2}$ and line width of 2 km~s$^{-1}$ for all three molecular species. The light blue shaded region shows the temperature range derived from the \ce{HCN}/\ce{HNC} ratio, with the blue line indicating the mean gas temperature. The gray shaded area shows the density ranged for the N-S ridge derived from envelope models.}
	\label{fig:highvslowratios}
\end{figure*}

Line ratio models were done using non-LTE slab calculation with RADEX \citep{vandertak2007} with molecular data files \citep{dumouchel2010,faure2007} obtained from the Leiden Atomic and Molecular Database \citep{schoier2005,vandertak2020}.
A line width of 2 km~s$^{-1}$ and column density of 10$^{13}$ cm$^{-1}$ were adopted.
The \ce{H2} density is constrained by the envelope models of \iras~\citep{crimier2010}, which indicate \ce{H2} densities below 10$^{6}$ cm$^{-3}$ at radius of 30 to 40$\arcsec$, equivalent to the edge of our Band 3 maps.
At radii below 10$\arcsec$, the \ce{H2} densities are between 10$^{6}$ and 10$^{7}$ cm$^{-3}$, increasing to $\sim$10$^{9}$ cm$^{-3}$ at the continuum peak of \iras~A.
Figure~\ref{fig:highvslowratios} shows the modeled ratios, with the temperature and density constraints obtained from the \ce{HCN}/\ce{HNC} map and envelope models, respectively.

Along the N-S ridge, \ce{HCN} 4--3/1--0 $\sim$ 0.02, and \ce{HNC} 4--3/1--0 $\sim$ 0.001.
These are consistent with the temperature and density constraints obtained from the \ce{HCN}/\ce{HNC} ratio and the envelope model.
Around the \iras~sources, the \ce{HCN} ratio is between 0.1 and 0.2. 
Considering the \ce{H2} densities from envelope models around these sources, above 10$^{8}$ cm$^{-3}$, the temperatures go well above 40 K, consistent with the \ce{HCN}/\ce{HNC}.

\section{Accretion rate estimates}
\label{app:accretion}

Assuming the north part of the N-S ridge is infalling, and thus delivering material onto source A, a rough estimate of the accretion rates can be done.
For the mass of the N-S ridge, a cylindrical volume can be assumed using a radius of 2.5$\arcsec$, and a length of 40$\arcsec$.
Based on the line ratio analysis, an \ce{H2} density of 10$^{6}$ cm$^{-3}$ is assumed. 
This results in a mass of 0.014 M$_{\odot}$. 
Free fall time scale was obtained by the equation t$_{ff}$ = ($R^3$ / $GM$)$^{1/2}$, where $R$ and $M$ are the enclosed radius and mass respectively, and $G$ is the gravitational constant.
Using $R$ = 20$\arcsec$ or 10$\arcsec$, and an enclosed mass of 2 M$_{\odot}$ \citep{maureira2020}, we get an accretion rate between 7$\times$10$^{-7}$ to 10$^{-6}$ M$_{\odot}$~yr$^{-1}$.
Changing the \ce{H2} density (0.5 to 5 $\times$ 10$^6$ cm$^{-3}$) or the enclosed mass (e.g., to 4 M$_{\odot}$) only changes the accretion rate by a few, but stays within the 10$^{-7}$ to 10$^{-6}$ M$_{\odot}$~yr$^{-1}$ range.

The above derived accretion rate needs to be compared to an accretion rate on source A.
The best estimate for this is from accretion luminosity $L_{acc}$.
Accretion rate $\dot{M}_{acc}$ is thus given by the relation $\dot{M}_{acc}$ = $L_{acc}$ / $RGM_{*}$, where $G$ is the gravitational constant, $M_{*}$ is the mass of the central source, and $R$ is the protostellar radius, assumed to be 3 $R_{\odot}$ \citep{dunham2010,hsieh2019}.
Assuming that $L_{bol}$ = $L_{acc}$ = 18 L$_{\odot}$ for source A and a combined central source mass of 2 M$_{\odot}$ \citep{maureira2020}, the accretion rate is about 8$\times$10$^{-7}$ M$_{\odot}$~yr$^{-1}$.
This accretion rate increases if only one of the components in the close binary is receiving material.
Making similar assumptions for source B ($L_{bol}$ = $L_{acc}$ = 3 L$_{\odot}$) yields an accretion rate of a few 10$^{-7}$ M$_{\odot}$~yr$^{-1}$.

Comparing the accretion rate from luminosity for source A to that derived for the N-S ridge further supports the scenario that the north part of the N-S ridge is likely an accretion flow delivering material.
Further high spectral resolution observations with a larger field of view are necessary to confirm this scenario.

\end{appendix}

\end{document}